\documentclass[apjl, twocolappendix]{emulateapj}

\pdfoutput=1
\usepackage[fleqn]{amsmath} % \dfrac, \gamma, ...
\usepackage{amsfonts} % \mathbb \forall
\usepackage{amstext}
\usepackage{amssymb} % ulteriori simboli (ex: \blacksquare)
\usepackage[colorlinks=true, citecolor=blue, linkcolor=blue, breaklinks=true]{hyperref}
\usepackage{natbib}
\newcommand{\hi } {{\rm H}\,{\small\rm I} \,}
\newcommand{\hiA} {{\rm H}\,{\small\rm I}}

\begin{document}
\nocite{*}
\title{One law to rule them all: the radial acceleration relation of galaxies}
\author{Federico Lelli$^{1, 2,\star}$}\thanks{$^{\star}$ESO Fellow; e-mail: flelli@eso.org}
\author{Stacy S. McGaugh$^{1}$} 
\author{James M. Schombert$^{3}$}
\author{Marcel S. Pawlowski$^{1,4,\dag}$}\thanks{$^{\dag}$Hubble Fellow}
\affil{$^{1}$Department of Astronomy, Case Western Reserve University, Cleveland, OH 44106, USA\\
$^{2}$European Southern Observatory, Karl-Schwarzschild-Strasse 2, D-85748, Garching, Germany\\
$^{3}$Department of Physics, University of Oregon, Eugene, OR 97403, USA\\
$^{4}$Department of Physics and Astronomy, University of California, Irvine, CA 92697, USA}
\begin{abstract}
We study the link between baryons and dark matter in 240 galaxies with spatially resolved kinematic data. Our sample spans 9 dex in stellar mass and includes all morphological types. We consider (i) 153 late-type galaxies (LTGs; spirals and irregulars) with gas rotation curves from the SPARC database; (ii) 25 early-type galaxies (ETGs; ellipticals and lenticulars) with stellar and \hi data from ATLAS$^{\rm 3D}$ or X$-$ray data from $Chandra$; and (iii) 62 dwarf spheroidals (dSphs) with individual-star spectroscopy. We find that LTGs, ETGs, and ``classical'' dSphs follow the same radial acceleration relation: the observed acceleration ($g_{\rm obs}$) correlates with that expected from the distribution of baryons ($g_{\rm bar}$) over 4 dex. The relation coincides with the 1:1 line (no dark matter) at high accelerations but systematically deviates from unity below a critical scale of $\sim$10$^{-10}$\,m\,s$^{-2}$. The observed scatter is remarkably small ($\lesssim$0.13 dex) and largely driven by observational uncertainties. The residuals do not correlate with any global or local galaxy property (baryonic mass, gas fraction, radius, etc.). The radial acceleration relation is tantamount to a Natural Law: when the baryonic contribution is measured, the rotation curve follows, and vice versa. Including ultrafaint dSphs, the relation may extend by another 2 dex and possibly flatten at $g_{\rm bar}\lesssim10^{-12}$\,m\,s$^{-2}$, but these data are significantly more uncertain. The radial acceleration relation subsumes and generalizes several well-known dynamical properties of galaxies, like the Tully-Fisher and Faber-Jackson relations, the ``baryon-halo'' conspiracies, and Renzo's rule.
\end{abstract}

\keywords{dark matter --- galaxies: dwarf --- galaxies: ellitpical and lenticular, cD --- galaxies: irregular --- galaxies: kinematics and dynamics --- galaxies: spiral}

\section{Introduction}\label{sec:intro}

The flat rotation curves of spiral galaxies \citep{Bosma1978, Rubin1978} provided clear empirical evidence for mass discrepancies in galactic systems, which are commonly attributed to non-baryonic dark matter (DM). Over the past 40 years, the relations between the baryonic and dynamical properties of galaxies have been intensively debated considering all different galaxy types: early-type galaxies (ETGs) like ellipticals and lenticulars \citep[e.g.,][]{Faber1976, Djorgovski1987, Dressler1987, Cappellari2016}, late-type galaxies (LTGs) like spirals and irregulars \citep[e.g.,][]{Rubin1985, vanAlbada1986, Persic1991, vanderKruit2011}, and dwarf spheroidals (dSphs) in the Local Group \citep[e.g.,][]{Mateo1998, Battaglia2008, Strigari2008, 2009ApJ...704.1274W}.

LTGs follow a tight baryonic Tully-Fisher \citep{Tully1977} relation (BTFR), linking the baryonic mass (stars plus gas) to the flat rotation velocity $V_{\rm f}$ \citep[e.g.,][]{McGaugh2000, Verheijen2001b, Lelli2016}. Similarly, ETGs follow the Faber-Jackson relation \citep{Faber1976}, linking stellar mass and stellar velocity dispersion $\sigma_{\star}$. \citet{denHeijer2015} show that ETGs with extended \hi disks (tracing $V_{\rm f}$) follow the same BTFR as LTGs. If $\sigma_{\star}$ is used to estimate $V_{\rm f}$, the most luminous dSphs adhere to the BTFR but the ultrafaint dSphs seem to deviate \citep{McGaugh2010}, possibly hinting at out-of-equilibrium dynamics due to tides from the host galaxy.

The BTFR is a ``global'' scaling relation between the baryonic and dynamical masses of galaxies. \citet{Sancisi2004} further advocates for a ``local'' Tully-Fisher kind of relation, linking baryons and dynamics on a radial basis. For LTGs, indeed, it has become clear that the rotation curve shape and the baryonic mass distribution are closely related \citep{Kent1987, Corradi1990, Casertano1991, Sancisi2004, McGaugh2004, McGaugh2005, Noordermeer2007a, Swaters2009, Swaters2012, Swaters2014, Lelli2010, Lelli2013, Lelli2016}. A similar coupling is also observed in ETGs \citep{Serra2016}.

Several works try to parametrize this baryon$-$DM coupling using different approaches \citep{Sanders1990, Persic1991, McGaugh1999, McGaugh2004, McGaugh2014, Swaters2009, Swaters2014, Lelli2013, Lelli2016b, Walker2014}. In particular, \citet{McGaugh2004} defines the ``mass discrepancy'' as $M_{\rm tot}/M_{\rm bar} \simeq V_{\rm obs}^{2}/V_{\rm bar}^{2}$ at every radius, where $V_{\rm obs}$ is the observed rotation curve and $V_{\rm bar}$ is the baryonic contribution from the distribution of stars and gas. \citet{McGaugh2004} finds that the mass discrepancy anticorrelates with the baryonic acceleration\footnote{In this paper the general term ``acceleration'' will always refer to the centripetal radial acceleration.} $V_{\rm bar}^{2}/R$ \citep[see also][]{Sanders1990}, leading to a mass discrepancy-acceleration relation (MDAR). The sample of \citet{McGaugh2004}, however, has heterogeneous surface photometry in different bands, hence one needs to pick a different stellar mass-to-light ratio ($\Upsilon_{\star}$) for each galaxy to compute $V_{\rm bar}$. \citet{McGaugh2004} shows that the scatter in the MDAR and BTFR can be simultaneously minimized by chosing an optimal $\Upsilon_{\star}$ for each galaxy, corresponding to the prescriptions of Modified Newtonian Dynamics \citep[MOND;][]{Milgrom1983}. This approach evinces the existence of a tight MDAR but does not enable measurement of its shape and scatter. Besides, the location of ETGs and dSphs on the MDAR was not addressed.

To improve the situation, we built the $Spitzer$ Photometry and Accurate Rotation Curves (SPARC) database: a sample of 175 disk galaxies (S0 to dIrr) with homogeneous [3.6] surface photometry and high-quality \hiA/H$\alpha$ rotation curves \citep[][hereafter Paper I]{Lelli2016b}. Several lines of evidence suggest that $\Upsilon_{\star}$ does not vary strongly at [3.6] \citep{McGaugh2014b, McGaugh2015, Meidt2014, Schombert2014a}, hence one can effectively use a single value of $\Upsilon_{\star}$ for disk galaxies of different masses and morphologies. This is a major improvement over previous studies. 

In \citet{McGaugh2016}, we provide a concise summary of our results from 153 SPARC galaxies. Here we present a more extensive analysis, including 25 ETGs and 62 dSphs. Specifically, we study the \textit{local} link between baryons and DM using a parametrization that minimizes observational uncertainties and degeneracies. We plot the baryonic acceleration ($g_{\rm bar} = V_{\rm bar}^2/R$) against the total acceleration ($g_{\rm obs} = V_{\rm obs}^2/R$) instead of the mass discrepancy ($V_{\rm obs}^{2}/V_{\rm bar}^{2}$). This has two key advantages: (i) the two axes are fully independent (photometry versus kinematics as in the BTFR), and (ii) the uncertainties on $\Upsilon_{\star}$ only enters in $g_{\rm bar}$. This parametrization is also used by \citet{Scarpa2006} for pressure-supported systems and by \citet{Wu2015} for rotation-supported galaxies using the best-fit data of \citet{McGaugh2004}. We find that $g_{\rm obs}$ correlates with $g_{\rm bar}$ over $\sim$4 dex ($\sim$6 dex including ultrafaint dSphs). This radial acceleration relation is remarkably tight: the observed scatter is only 0.13 dex and largely driven by observational uncertainties.

We start by describing our galaxy samples and associated data (Sect.\,\ref{sec:Data}). Next, we study the radial acceleration relation of LTGs (Sect.\,\ref{sec:LTGs}), test different normalizations of $\Upsilon_{\star}$ (Sect.\,\ref{sec:ML}), and explore different choices of the dependent $x$ variable (Sect.\,\ref{sec:Stellar}). We then investigate the same relation for ETGs (Sect.\,\ref{sec:ETGs}) and dSphs (Sect.\,\ref{sec:dSphs}). Finally, we discuss the link with other dynamical laws of galaxies (Sect.\,\ref{sec:Other}) and the implications for galaxy formation models and alternative theories (Sect.\,\ref{sec:Impl}).

\begin{figure}[thb]
\centering
\includegraphics[width=0.47\textwidth]{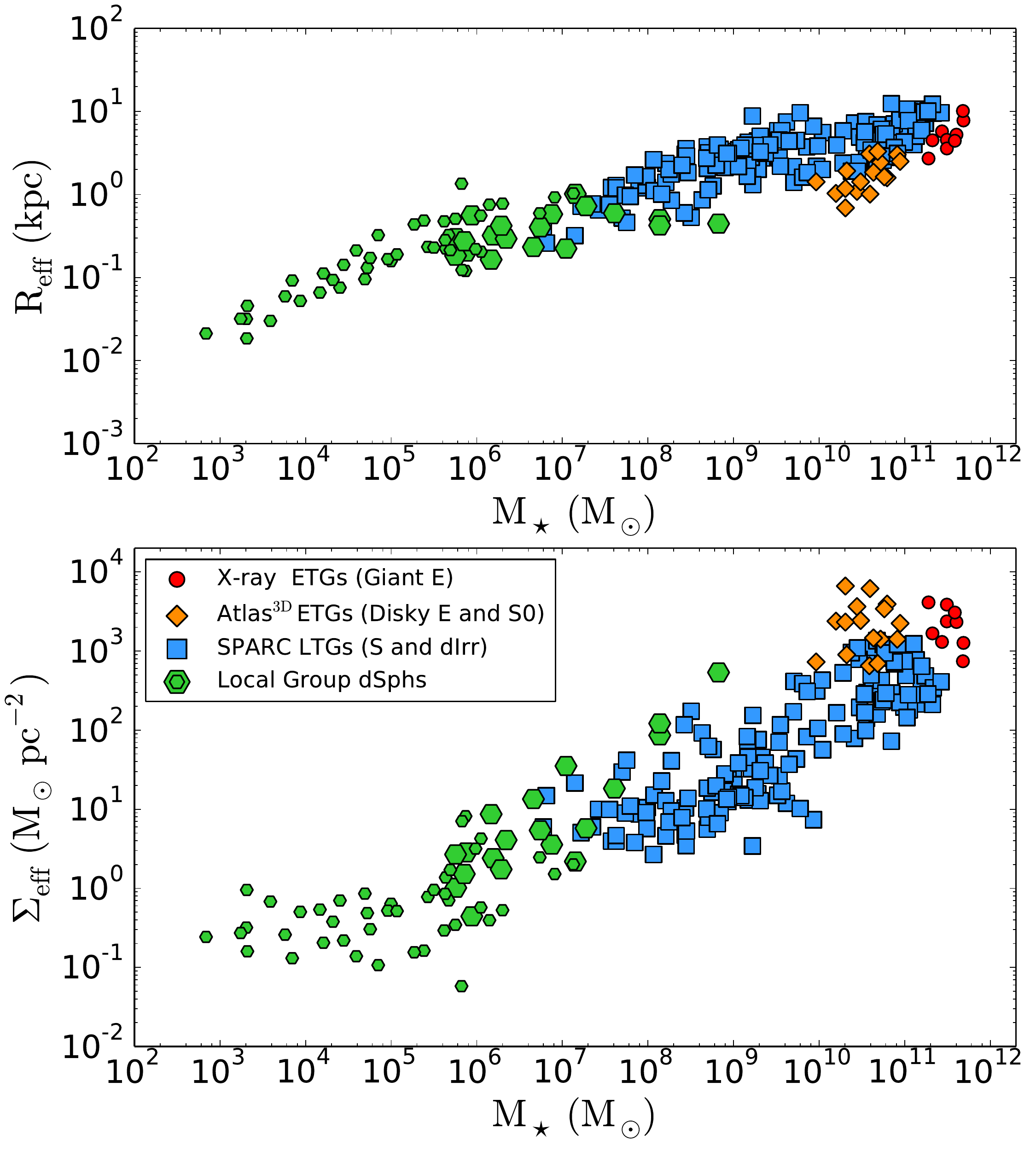}
\caption{Structural properties of the 240 galaxies in our sample. The stellar mass is plotted against the effective radius (top) and effective surface density (bottom). Different symbols indicate different galaxy types as given by the legend. For dSphs, large symbols show the classical satellites of the MW and M31, while small symbols indicate ultrafaint dSphs and more isolated dSphs (like Tucana and Cetus). Stellar masses and effective surface densities are computed using the mass-to-light ratios in Table\,\ref{tab:ML}.}
\label{fig:Sample}
\end{figure}
\section{Galaxy Sample and Data Analysis}\label{sec:Data}

Our sample comprises 240 galaxies of all main morphological types. Figure\,\ref{fig:Sample} shows their structural properties, using Spitzer photometry for LTGs (Paper\,I) and ETGs (Sect.\,\ref{sec:ETGData}) and $V$-band photometry for dSphs (Sect.\,\ref{sec:dSphData}). Our sample covers the widest possible range of galaxy properties, spanning 9 dex in stellar mass, 5 dex in effective surface density, and 3 dex in effective radius\footnote{The effective radius $R_{\rm eff}$ is defined as the geometrical radius $R=\sqrt{ab}$ that encompasses half of the total luminosity. The effective surface density $\Sigma_{\rm eff}$ is then simply $(\Upsilon_{\star}L)/(2\pi R_{\rm eff}^2)$.}. The cold gas fractions range from $\sim$90$\%$ in dIrrs to virtually zero in dSphs. Here we describe photometric and kinematic data for our sub-samples of 153 LTGs, 25 ETGs, and 62 dSphs.

\subsection{Late-Type Galaxies}\label{sec:DiskData}

We consider 153 galaxies from the SPARC database, providing [3.6] surface brightness profiles, \hiA/H$\alpha$ rotation curves, and mass models (Paper I). SPARC spans a wide range of disk properties and contains representatives of all late-type morphologies, from bulge-dominated spirals to gas-dominated dwarf irregulars. Three SPARC galaxies are classified as lenticulars by \citet{RC3}: we consider them among the LTGs to have a clear separation from the S0s of Atlas$^{\rm 3D}$ \citep{Cappellari2011}, having different types of data (Sect.\,\ref{sec:ETGData}).

Paper\,I describes the analysis of [3.6] images, the collection of rotation curves, and the derivation of mass models. SPARC contains a total of 175 galaxies, but we exclude 12 objects where the rotation curve may not trace the equilibrium gravitational potential (quality flag $Q=3$) and 10 face-on galaxies ($i < 30^{\circ}$) with uncertain velocities due to large $\sin(i)$ corrections (see Paper\,I). This does not introduce any selection bias because galaxy disks are randomly oriented on the sky.

\subsubsection{The total gravitational field of LTGs}

LTGs possess a dynamically-cold \hi disk, hence the observed rotation curve $V_{\rm obs}(R)$ directly traces the gravitational potential at every radii. Corrections for pressure support are relevant only in the smallest dwarf galaxies with $V_{\rm obs}\simeq 20$ km~s$^{-1}$ \citep[e.g.,][]{Lelli2012b}. The total gravitational field is given by
\begin{equation}
 g_{\rm obs}(R) = \dfrac{V_{\rm obs}^{2}(R)}{R} = -\nabla \Phi_{\rm tot}(R),
\end{equation}
where $\phi_{\rm tot}$ is the total potential (baryons and DM). 

The uncertainty on $g_{\rm obs}$ is estimated as
\begin{equation}
 \delta_{g_{\rm obs}} = g_{\rm obs} 
 \sqrt{\bigg[\dfrac{2\delta_{V_{\rm obs}}}{V_{\rm obs}}\bigg]^{2} 
 + \bigg[\dfrac{2\delta_{i}}{\tan(i)}\bigg]^{2} 
 + \bigg[\dfrac{\delta_{D}}{D}\bigg]^{2}}.
\end{equation}
The errors on the rotation velocity ($\delta_{V_{\rm obs}}$), disk inclination ($\delta_{i}$), and galaxy distance ($\delta_{D}$) are described in Paper I. Here we consider only velocity points with $\delta_{V_{\rm obs}}/V_{\rm obs} < 0.1$. A minimum accuracy of 10$\%$ ensures that $g_{\rm obs}$ is not affected by strong non-circular motions or kinematic asymmetries. This criterion removes only $\sim$15$\%$ of our data (from 3149 to 2693 points), which are primarily in the innermost or outermost galaxy regions. Dropping this criterion does not change our results: it merely increases the observed scatter as expected from less accurate data. The mean error on $g_{\rm obs}$ is 0.1 dex.

\subsubsection{The baryonic gravitational field of LTGs}

We compute the gravitational field in the galaxy midplane by numerically solving Poisson's equation, using the observed radial density profiles of gas and stars and assuming a nominal disk thickness (see Paper\,I):
\begin{equation}\label{Eq:gbar}
 g_{\rm bar}(R) = \dfrac{V_{\rm bar}^{2}(R)}{R} = -\nabla \Phi_{\rm bar}(R).
\end{equation}
Specifically, the expected velocity $V_{\rm bar}$ is given by
\begin{equation}
 V_{\rm bar}^2(R) = V_{\rm gas}^2(R) + \Upsilon_{\rm disk}V_{\rm disk}^2(R) + \Upsilon_{\rm bul}V_{\rm bul}^2(R),
\end{equation}
where $\Upsilon_{\rm disk}$ and $\Upsilon_{\rm bul}$ are the stellar mass-to-light ratios of disk and bulge, respectively (see Sect.\,\ref{sec:MLsp}).

The uncertainty on $g_{\rm bar}$ is estimated considering a typical 10$\%$ error on the \hi flux calibration and a 25$\%$ scatter on $\Upsilon_{\star}$. The latter is motivated by SPS models \citep{McGaugh2014, Meidt2014, Schombert2014a} and the BTFR \citep{McGaugh2015, Lelli2016}. Uncertainties in the $Spitzer$ photometric calibration are negligible. Note that $g_{\rm bar}$ is distance independent because both $V_{\rm bar}^{2}$ and $R$ linearly vary with $D$. The mean error on $g_{\rm bar}$ is 0.08 dex but this is a lower limit because it neglects uncertainties due to the adopted 3D geometry (vertical disk structure and bulge flattening) and the possible (minor) contribution of molecular gas (see Paper I for details). This may add another $\sim$20$\%$ uncertainty to points at small radii.

\subsection{Early-Type Galaxies}\label{sec:ETGData}

ETGs generally lack high-density \hi disks, thus one cannot derive gas rotation curves that directly trace $\Phi_{\rm tot}$ at every radius. Different tracers, however, have been extensively explored. Integral field units (IFUs) have made it possible to study the 2D stellar kinematics of ETGs, revealing that $\sim$85$\%$ of them are rotating \citep[see][for a review]{Cappellari2016}. X$-$ray telescopes like \emph{Chandra} and \emph{XMM} have been used to probe the hot gas and trace total mass profiles assuming hydrostatic equilibrium \citep[see][for a review]{Buote2012}. Progess has also been made using discrete kinematical tracers (planetary nebule and globular clusters) and strong gravitational lensing \citep[see][for a review]{Gerhard2013}.

Here we consider two different data sets of ETGs: (i) 16 ETGs from Atlas$^{\rm 3D}$ \citep{Cappellari2011} that have inner rotating stellar components \citep{Emsellem2011} and outer low-density \hi disks/rings \citep{denHeijer2015}; and (ii) 9 ETGs with relaxed X$-$ray haloes. These two data sets roughly correspond to two different ``families'' of ETGs \citep[e.g.,][]{Kormendy1996, Kormendy2009, Cappellari2016}: rotating ETGs from Atlas$^{\rm 3D}$ are either lenticulars or disky ellipticals, while X$-$ray ETGs are generally giant boxy ellipticals. Three X$-$ray ETGs are part of Atlas$^{\rm 3D}$: NGC~4261 and NGC~4472 are classified as ``slow rotators'' (essentially non-rotating galaxies), while NGC~4649 is classified as a ``fast rotator'' but actually lies at the boundary of the classification ($V_{\rm rot}/\sigma_{\star}\simeq 0.12$). These X$-$ray ETGs are representative of ``classic'' pressure-supported ellipticals, possibly metal rich, anisotropic, and mildy triaxial. Appendix\,\ref{app:ETG} provides the basic properties of these 25 ETGs.

\subsubsection{The total gravitational field of rotating ETGs}\label{sec:rotETGs}

Our sample of rotating ETGs is drawn from \citet{Serra2016}. Despite the significant amount of rotation, the stellar components of these ETGs are dynamically hot ($V_{\rm rot}/\sigma_{\star} \lesssim 1$), thus pressure support needs to be taken into account. \citet{Cappellari2013} build Jeans anisotropic models (JAM), fitting the stellar velocity field, the velocity dispersion map, and the $r$-band image of the galaxy. These models assume that (i) the galaxy is axisymmetric and the velocity ellipsoid is nearly oblate ($\sigma_{\phi}\simeq\sigma_{\rm R}\gtrsim\sigma_{\rm z}$), (ii) the total mass distribution follows the light distribution, and (iii) the $r$-band mass-to-light ratio is constant with radius. The JAM models return the maximum circular velocity $V_{\rm max}$ within the IFU field of view (typically 1 effective radius). We use this quantity to estimate $g_{\rm JAM} = V_{\rm max}^2/R_{\rm max}$. We assume a formal error of 10$\%$ on $V_{\rm max}$ (similar to typical \hi data), but we stress that $V_{\rm max}$ is not directly observed and remains somewhat model dependent.

The situation is much simpler at large radii because \hi disks/rings directly trace the gravitational potential. \citet{denHeijer2015} analyse \hi velocity fields and position-velocity diagrams for these 16 ETGs, providing a velocity measurement in the outer parts ($V_{\hi}$). Hence, we have $g_{\rm \hi} = V_{\hi}^2/R_{\hi}$. In principle, high-resolution \hi observations may trace the full rotation curve in the outer parts of these ETGs, but we are currently limited to a single, average value. Therefore, we have ``rotation curves'' with two points: an inner point from IFU data (via JAM models) and an outer point from \hi data.

For NGC~2974, we note a small discrepancy between the values of $V_{\rm max}$ and $V_{\hi}$ reported by \citet{Serra2016} and \citet{Weijmans2008}. The latter is an in-depth study providing the full stellar, H$\alpha$, and \hi rotation curves. We adopt the values from \citet{Weijmans2008}.

\subsubsection{The total gravitational field of X$-$ray ETGs}\label{sec:XrayETGs}

We consider 9 ETGs from \citet{Humphrey2006, Humphrey2008, Humphrey2009, Humphrey2011, Humphrey2012}. This series of papers provides a homegeneous analysis of deep X$-$ray observations from \emph{Chandra} and \emph{XMM}. The X$-$ray data give accurate density and temperature profiles, hence one can directly compute the enclosed total mass $M_{\rm tot}(<r)$ using the equation of hydrostatic equilibrium \citep[see][for a review]{Buote2012}. The main assumptions are (i) the hot gas is in hydrostatic equilibrium, (ii) the system is spherically symmetric, and (iii) the thermal pressure dominates over non-thermal components like magnetic pressure or turbulence. For our 9 ETGs, these assumptions are realistic \citep{Buote2012}. 

We consider total mass profiles from the ``classic'' smoothed inversion approach \citep[see][]{Buote2012}. This technique is non-parametric (no specific potential is assumed a priori) and provides the enclosed mass at specific radii. The sampling is determined by the quality of the data, i.e., by ``smoothing`` the temperature and density profiles to ensure that the resulting mass profile monotonically increases with radius. This technique is closest to the derivation of rotation curves in LTGs. The ``observed acceleration'' is then given by
\begin{equation}\label{Eq:dSph2}
 g_{\rm obs}(r) = -\nabla\Phi_{\rm tot}(r) = \dfrac{G M_{\rm tot}(<r)}{r^2},
\end{equation}
where $G$ is Newton constant and $r$ is the 3D radius.

Four of these ETGs (NGC\,1407, NGC\,4261, NGC\,4472, and NGC\,4649) belong to groups with 10 to 60 known members, thus the outer mass profile may be tracing the group potential instead of the galaxy potential \citep{Humphrey2006}. For these galaxies, we restrict our analysis to $R<4 R_{\rm eff}$. This radius is determined by computing rotation curves from the mass profiles. For example, NGC\,4261 has a relatively flat rotation curve within $4 R_{\rm eff} \simeq 18$ kpc but starts to rise at larger radii, reaching velocities of $\sim$700 km~s$^{-1}$ at 100 kpc. Clearly, such rotation curve cannot be due to the galaxy potential. The other five ETGs are relatively isolated, hence we use the full mass profile out to $\sim$15 $R_{\rm eff}$.

We also exclude data at $R < 0.1 R_{\rm eff}$ for three basic reasons: (i) intermittent feedback from active galactic nuclei may introduce significant turbulence, breaking the assumption of hydrostatic equilibrium at small radii \citep{Buote2012}, (ii) the gravitational effect of central black holes may be important \citep{Humphrey2008, Humphrey2009}, but these dark components are not included in our baryonic mass models (Sect.\,\ref{sec:barETGs}), and (iii) X$-$ray ETGs generally have steep luminosity profiles (Sersic index $\gtrsim$ 4, e.g., \citealt[][]{Kormendy2009}) which may be smeared by the limited spatial resolution of $Spitzer$, leading to inaccurate mass models in the innermost parts. This quality criteria retains 80 out of 97 points.

\subsubsection{The baryonic gravitational field of ETGs}\label{sec:barETGs}

For consistency with SPARC, we collect [3.6] images from the $Spitzer$ archive and derive surface photometry using the same procedures as in Paper\,I. For two objects (NGC\,1521 and NGC\,6798), $Spitzer$ images are not available: we use WISE images at 3.2 $\mu$m (W1 band) and convert W1 profiles to [3.6] adopting W1-[3.6]=0.11. For these two ETGs, we do not consider the innermost regions ($R<6'' \simeq 4$ pixels) due to the limited spatial resolution of WISE. The surface brightness profiles are used to calculate $g_{\rm bar}$ (Eq.\,\ref{Eq:gbar}). Stellar mass-to-light ratios are discussed in Sect.\,\ref{sec:MLsp}. We neglect the contribution of cold gas (\hi or H$_2$) as this is a minor dynamical component for ETGs and can be safely ignored \citep{Weijmans2008}. For X$-$ray ETGs, the contribution of hot gas is calculated assuming spherical symmetry and using published density profiles (see Appendix\,\ref{app:ETG} for references).

Among our rotating ETGs, 12/16 galaxies are morphologically classified as lenticulars, while the remaining 4 ellipticals may have face-on disks as revealed by their stellar kinematics \citep[see e.g.,][]{Cappellari2016}. We perform non-parametric bulge-disk decompositions using the same strategy as in Paper\,I. In particular, structures in the luminosity profiles due to bars or lenses are assigned to the disk. We assume that bulges are spherical, while disks have an exponential vertical distribution with scale-height $z_{\rm d} = 0.196 R_{\rm d}$ (see Paper\,I).

All X$-$ray ETGs are classified as ellipticals with the exceptions of NGC~1332 (S0). For this galaxy, we perform a bulge-disk decomposition, while the other X$-$ray ETGs are treated as spherically symmetric systems.

\subsection{Dwarf Spheroidals}\label{sec:dSphData}

For every currently known dSph in the Local Group, we collected distances, $V$-band magnitudes, half-light radii, ellipticities, and velocity dispersions from the literature. We only exclude Sagittarius and Bootes\,III which are heavily disrupted satellites of the Milky Way \citep[MW,][]{Ibata1994, Carlin2009}. In Appendix \ref{sec:Catalogue}, we describe the entries of this catalogue and provide corresponding references. Here we stress that velocity dispersions ($\sigma_{\star}$) are derived from high-resolution individual-star spectroscopy and their errors are indicative. The reliability of these measurements depends on the available number of stars, the possible contaminations from foreground objects and binary stars, and the dynamical state of the galaxy \citep[e.g.,][]{Walker2009b, McConnachie2010, McGaugh2010, Minor2010}. 

In general, the data quality for ``classical'' dSphs and ``ultrafaint'' dSphs is markedly different. Classical dSphs were discovered during the 20$^{\rm th}$ century and their properties have been steadily refined over the years. Ultrafaint dSphs have been discovered during the past $\sim$10 years with the advent of the Sloan Digital Sky Survey \citep[e.g.,][]{Willman2005} and Dark Energy Survey \citep[e.g.,][]{2015ApJ...805..130K}. Their properties remain uncertain due to their extreme nature: some ultrafaint dSphs are less luminous than a single giant star! In our opinion, it is still unclear whether all these objects deserve the status of ``galaxies'', or whether they are merely overdensities in the stellar halo of the host galaxy. Hence, caution is needed to interpret the data of ultrafaint dSphs.

\begin{figure*}[thb]
\centering
\includegraphics[width=0.95\textwidth]{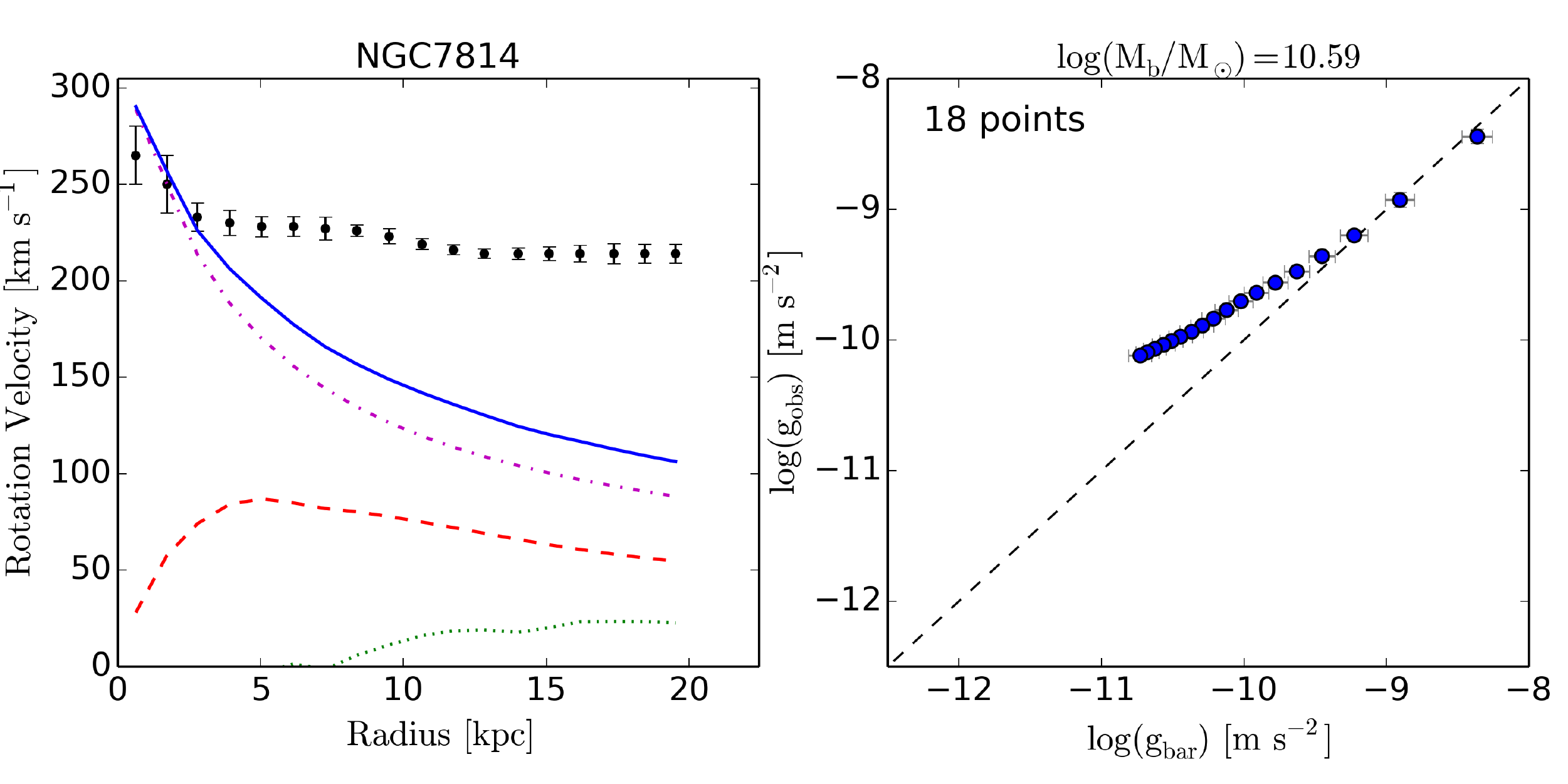}
\caption{A movie shows how each individual SPARC galaxy adds to the radial acceleration relation. The galaxy name and its total baryonic mass are indicated to the top. The left panel shows the observed rotation curve (dots) and the expected baryonic contributions: gas (dotted line), stellar disk (dashed line), bulge (dash-dotted line), and total baryons (solid line). The right panel shows the galaxy location on the $g_{\rm obs}$-$g_{\rm bar}$ plane; the cumulative amount of datapoints is indicated to the top left corner. The radial acceleration relation (right panel) evinces a remarkable regularity between baryons and dynamics, in spite of the large diversity in individual mass models (left panel). This movie is also available on the SPARC webpage (\href{url}{astroweb.cwru.edu/SPARC}) in different formats.}
\label{fig:Video}
\end{figure*}
\subsubsection{The total gravitational field of dSphs}

Dwarf spheroidals typically lack a rotating gas disk and have a fully pressure-supported stellar component. The gravitational field cannot be directly estimated at every radius due to degeneracies between the projected velocity dispersion ($\sigma_{\star}$) and the velocity dispersion anisotropy. Several studies, however, show that the total mass within the half-light radius (hence, $g_{\rm obs}$) does not strongly depend on anisotropy and can be estimated from the average $\sigma_{\star}$ \citep{2009ApJ...704.1274W, Wolf2010, Amorisco2011}. Following \citet{Wolf2010}, we have
\begin{equation}\label{Eq:dSph1}
 g_{\rm obs}(r_{1/2}) = -\nabla\Phi_{\rm tot}(r_{1/2}) = 3 \dfrac{\sigma_{\star}^{2}}{r_{1/2}},
\end{equation}
where $r_{1/2}$ is the deprojected 3D half-light radius. Eq.\,\ref{Eq:dSph1} assumes that (i) the system is spherically symmetric, (ii) the projected velocity dispersion profile is fairly flat near $r_{1/2}$, and (iii) $\sigma_{\star}$ traces the equilibrium potential. These assumptions are sensible for classical dSphs, but remain dubious for many ultrafaint dSphs \citep{McGaugh2010}. The error on $g_{\rm obs}$ is estimated considering \textit{formal} errors on $\sigma_{\star}$ and $r_{1/2}$ (including distance uncertainties). 

We note that the brightest satellites of M31 (NGC\,147, NGC\,185, and NGC\,205) are sometimes classified as dwarf ellipticals (dEs) and their stellar components show significant rotation at large radii \citep{2010ApJ...711..361G}, similar to dEs in galaxy clusters \citep[e.g.,][]{vanZee2004}. At $r < r_{1/2}$, however, the stellar velocity dispersion of these galaxies is much larger than the stellar rotation, hence Eq.\,\ref{Eq:dSph1} still provides a sensible estimate of $g_{\rm obs}$.

\begin{table}
\begin{center}
\caption{Fiducial stellar mass-to-light ratios}
\begin{tabular}{lc}
\hline
\hline
Galaxy Component & $\Upsilon_{\star}$\\
\hline
Disks of LTGs  (Sa to Irr)    & 0.5 $M_{\odot}/L_{\odot}$\\
Bulges of LTGs (Sa to Sb)     & 0.7 $M_{\odot}/L_{\odot}$\\
Rotating ETGs (S0 and disky E)& 0.8 $M_{\odot}/L_{\odot}$\\
X$-$ray ETGs   (Giant metal-rich E)  & 0.9 $M_{\odot}/L_{\odot}$\\
Dwarf Spheroidals ($V$-band) & 2.0 $M_{\odot}/L_{\odot}$\\
\hline
\end{tabular}
\tablecomments{The values of $\Upsilon_{\star}$ refer to 3.6 $\mu$m apart for dwarf spheroidals ($V$-band). We adopt a \citet{Chabrier2003} IMF and the SPS models of \citet{Schombert2014a}.}
\label{tab:ML}
\end{center}
\end{table}
\subsubsection{The baryonic gravitational field of dSphs}

The baryonic gravitational field at $r_{1/2}$ is given by
\begin{equation}\label{Eq:dSph2}
 g_{\rm bar}(r_{1/2}) = -\nabla\Phi_{\rm bar}(r_{1/2}) = G \dfrac{\Upsilon_{\rm V} L_{\rm V}}{2 r_{1/2}^2},
\end{equation}
where $\Upsilon_{\rm V}$ is the $V$-band stellar mass-to-light ratio. We assume $\Upsilon_{\rm V} = 2$ $M_{\odot}$/$L_{\odot}$ as suggested by studies of resolved stellar populations. For example, \citet{deBoer2012a, deBoer2012b} derive accurate SFHs for Fornax and Sculptor using deep color-magnitude diagrams and assuming a \citet{Kroupa2001} IMF, which is very similar to the \citet{Chabrier2003} IMF. The inferred stellar masses imply $\Upsilon_{\rm V} = 1.7$ $M_{\odot}$/$L_{\odot}$ for Fornax and $\Upsilon_{\rm V} = 2.4$ $M_{\odot}$/$L_{\odot}$ for Sculptor, adopting a gas-recycling efficiency of 30$\%$. The error on $g_{\rm bar}$ is estimated considering a 25$\%$ scatter on $\Upsilon_{\rm V}$ (2.0$\pm$0.5 $M_{\odot}$/$L_{\odot}$) and formal errors on $L_{\rm V}$ and $r_{1/2}$.

\subsection{Stellar mass-to-light ratios at 3.6 $\mu$m}\label{sec:MLsp}
 
The stellar mass-to-light ratio at 3.6 $\mu$m is known to show a smaller scatter and weaker dependence on color than in optical bands \citep[e.g.,][]{McGaugh2014b, Meidt2014}. When comparing ETGs and LTGs, however, it is sensible to distinguish between different structural components and galaxy types, given their different metallicities and star formation histories (SFHs). Here we consider four different cases: (i) the star-forming disks of LTGs, (ii) the bulges of LTGs, (iii) the bulges and disks of rotating ETGs (S0 and disky E), and (iv) X$-$ray ETGs (massive and metal rich). We use the stellar population synthesis (SPS) models of \citet{Schombert2014a}, which assume a \citet{Chabrier2003} initial mass function (IMF) and include metallicity evolution. We consider constant SFHs for the disks of LTGs and exponentially declining SFHs for the other cases, exploring different timescales and metallicities. Our fiducial values are summarized in Table\,\ref{tab:ML}. These values are in good agreement with different SPS models \citep{McGaugh2014b, Meidt2014, Norris2016}.

For LTGs, our values of $\Upsilon_{\star}$ agrees with resolved stellar populations in the LMC \citep{Eskew2012}, provides sensible gas fractions (Paper I), and minimizes the scatter in the BTFR \citep{Lelli2016}. The BTFR scatter is very small for a fixed value of $\Upsilon_{\star}$ at [3.6], hence the actual $\Upsilon_{\star}$ cannot vary wildy among different galaxies. The role of $\Upsilon_{\star}$ is further investigated in Section\,\ref{sec:ML}.

\begin{figure*}[thb]
\centering
\includegraphics[width=0.95\textwidth]{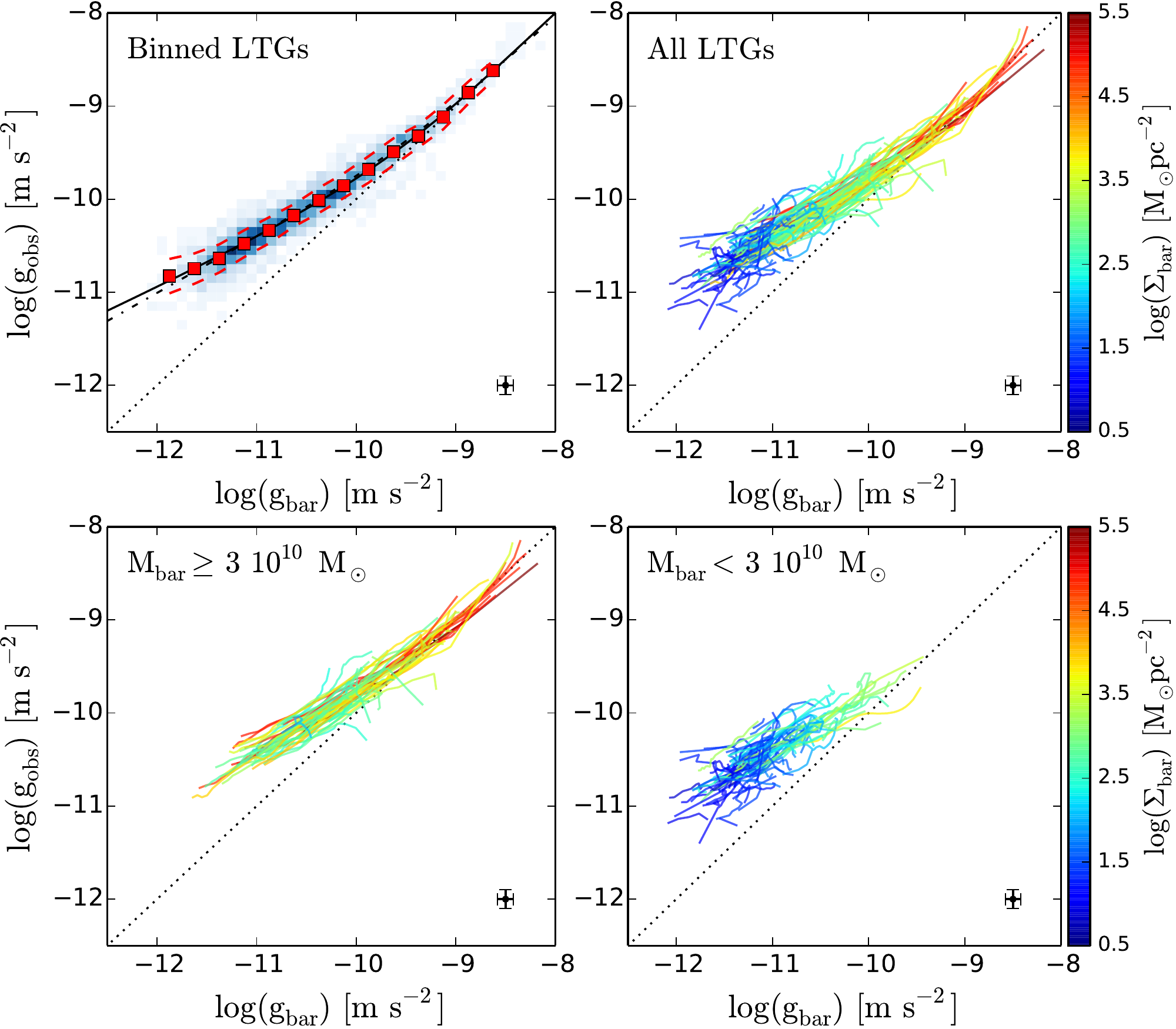}
\caption{Radial acceleration relation for LTGs. The total gravitational field ($g_{\rm obs}$) is derived at every radius from the rotation curve, while the baryonic gravitational field ($g_{\rm bar}$) is calculated from the distribution of stars and gas. Top left: The blue colorscale represents 2693 independent datapoints. Red squares and dashed lines show the mean and standard deviation of binned data, respectively. The dotted line indicates the 1:1 relation. Dash-dotted and solid lines show error-weighted fits using equations \ref{Eq:DoublePower} and \ref{eq:StacyFnc}, respectively. Top right: Each galaxy is plotted by a solid line, color-coded by the effective baryonic surface density. Bottom: Same as the top-right panel but we separate high-mass (left) and low-mass (right) galaxies. In all panels, the small dot to the bottom-right corner shows the typical errorbars on individual datapoints.}
\label{fig:RFR}
\end{figure*}
\section{Radial Acceleration Relation for Late-Type Galaxies}\label{sec:LTGs}

\subsection{General Results for LTGs}\label{sec:DiskRes}

Figure \ref{fig:Video} displays mass models for each SPARC galaxy (left) and their location on the $g_{\rm obs}-g_{\rm bar}$ plane (right). Figure \ref{fig:RFR} (top left) shows the resulting radial acceleration relation for our 153 LTGs. Since we have $\sim$2700 independent points, we use a 2D heat map (blue colorscale) and bin the data in $g_{\rm bar}$ (red squares). The radial acceleration relation is remarkably tight. Note that $g_{\rm obs}$ and $g_{\rm bar}$ are fully independent.

Figure \ref{fig:RFR} (top right) show the radial acceleration relation using a solid line for each LTG. In this plot one cannot appreciate the actual tightness of the relation because many lines fall on top of each other and a few outliers become prominent to the eye. This visualization, however, illustrates how different galaxies cover different regions of the relation. We color-code each galaxy by the equivalent baryonic surface density \citep{McGaugh2005}:
\begin{equation}
 \Sigma_{\rm bar} = \dfrac{3}{4}\dfrac{M_{\rm bar}}{R_{\rm bar}^{2}},
\end{equation}
where $M_{\rm bar} = M_{\rm gas} + M_{\star}$ is the total baryonic mass and $R_{\rm bar}$ is the radius where $V_{\rm bar}$ is maximum. 
In the bottom panels of Figure \ref{fig:RFR}, we also separate high-mass and low-mass galaxies at $M_{\rm bar} \simeq 3\times 10^{10}$ $M_{\odot}$. This value is similar to the characteristic mass in the $M_{\star}-M_{\rm halo}$ relation from abundance matching \citep{Moster2013} and may correspond to the transition between cold and hot modes of gas accretion \citep{Dekel2006}.

\begin{figure*}[thb]
\centering
\includegraphics[width=0.975\textwidth]{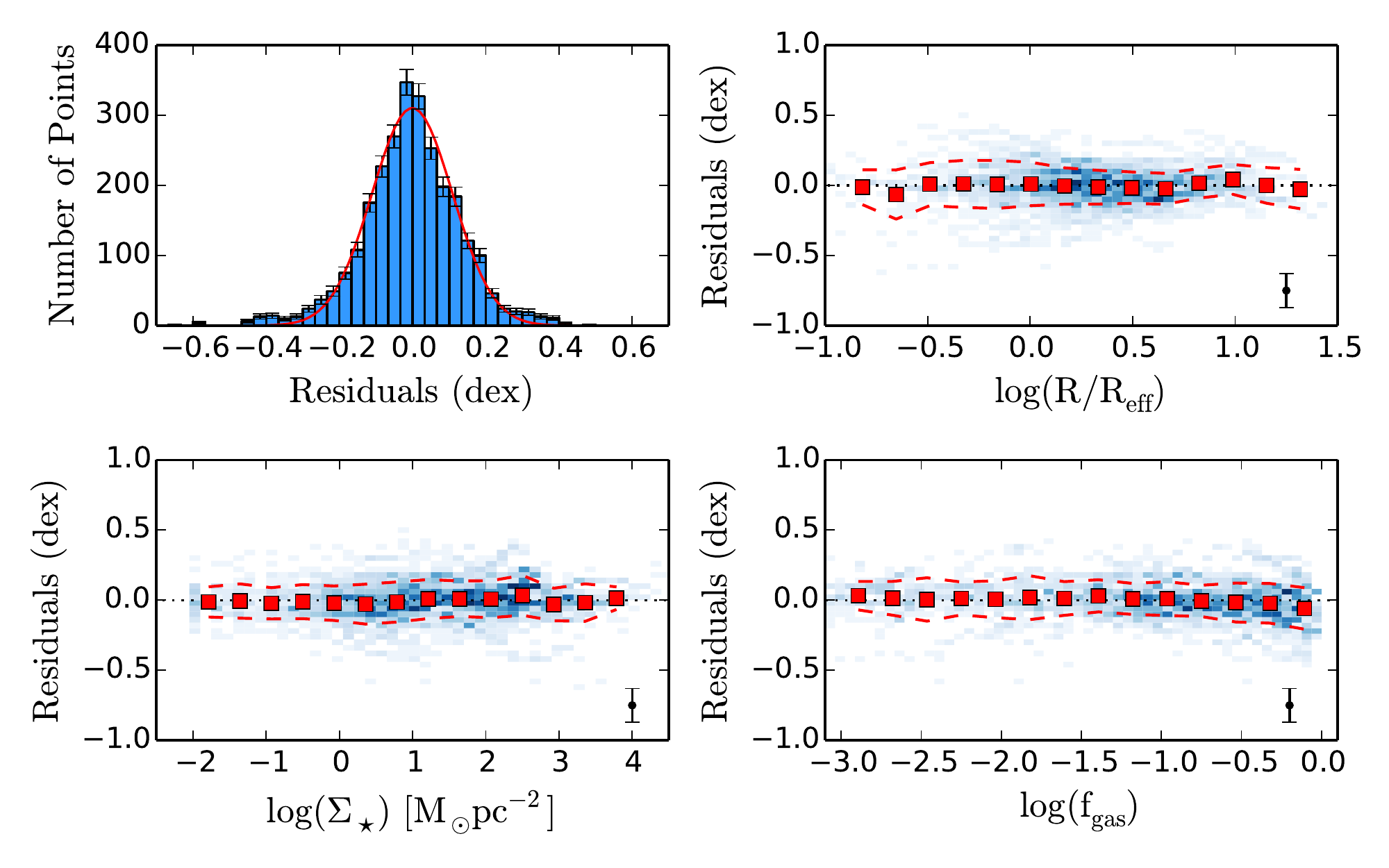}
\caption{Residuals after subtracting Eq.~\ref{eq:StacyFnc} from the radial acceleration relation. The top left panel shows an histogram with Poissonian ($\sqrt{N}$) errorbars and a Gaussian fit. The other panels show the residuals against radius (top right), stellar surface density at $R$ (bottom left), and local gas fraction $M_{\rm gas}/M_{\rm bar}\simeq V_{\rm gas}^2/V_{\rm bar}^2$ at $R$ (bottom right). Squares and dashed lines show the mean and standard deviation of binned residuals, respectively. The bar to the bottom-right corner shows the anticipated scatter from observational uncertainties.}
\label{fig:RFRres}
\end{figure*}
\begin{table*}
\begin{center}
%\resizebox{9cm}{!}{
%\setlength{\tabcolsep}{4pt}
\caption{Fits to the radial acceleration relation of LTGs using a double power law (Eq.\,\ref{Eq:DoublePower})}
\begin{tabular}{lccccccccc}
\hline
\hline
$x$ variable & $x$ and $\hat{x}$ units & $\Upsilon_{\rm disk}$ & $\Upsilon_{\rm bul}$ & $\alpha$ & $\beta$ & $\hat{x}$ & $\hat{y}$ & scatter (dex) \\
\hline
$g_{\rm bar}$      & 10$^{-10}$ m s$^{-2}$ & 0.5 & 0.7 & 0.94$\pm$0.06 & 0.60$\pm$0.01 & 2.3$\pm$1.1   & 2.6$\pm$0.8   & 0.13 \\
$g_{\rm bar}$      & 10$^{-10}$ m s$^{-2}$ & 0.5 & 0.5 & 1.03$\pm$0.07 & 0.60$\pm$0.01 & 2.3$\pm$0.6   & 1.8$\pm$0.7   & 0.14 \\
$g_{\rm bar}$      & 10$^{-10}$ m s$^{-2}$ & 0.2 & 0.2 & 1.59$\pm$1.32 & 0.77$\pm$0.01 & 10$\pm$22     & 22$\pm$37     & 0.15 \\
$g_{\star}$        & 10$^{-10}$ m s$^{-2}$ & 0.5 & 0.7 & 0.79$\pm$0.01 & 0.24$\pm$0.01 & 0.16$\pm$0.02 & 0.40$\pm$0.03 & 0.13 \\
$\Sigma_{\star}$   & $M_{\odot}$ pc$^{-2}$ & 0.5 & 0.7 & 0.64$\pm$0.03 & 0.09$\pm$0.01 & 9.3$\pm$2.1   & 0.48$\pm$0.03 & 0.25 \\
$\Sigma_{\rm disk}$& $M_{\odot}$ pc$^{-2}$ & 0.5 & 0.0 & 0.56$\pm$0.02 & 0.08$\pm$0.01 & 6.8$\pm$1.4   & 0.45$\pm$0.03 & 0.23 \\
\hline
\end{tabular}
\tablecomments{In all cases the $y$ variable is the observed radial acceleration $g_{\rm obs} = V_{\rm obs}^2/R$ and the units of $\hat{y}$ are 10$^{-10}$ m s$^{-2}$.}
\label{tab:Fit}
\end{center}
\end{table*}

High-mass, high-surface-brightness (HSB) galaxies cover the high-acceleration portion of the relation: they are baryon dominated in the inner parts and become DM dominated in the outer regions. Low-mass, low-surface-brightness (LSB) galaxies cover the low-acceleration portion: the DM content is already significant at small radii and systematically increase with radius. Strikingly, in the central portion of the relation ($10^{-11} \lesssim g_{\rm bar} \lesssim 10^{-9}$ cm s$^{-2}$) the inner radii of low-mass galaxies overlap with the outer radii of high-mass ones. \textit{The rotating gas in the inner regions of low-mass galaxies seems to relate to that in the outer regions of high-mass galaxies.}

\begin{figure*}[thb]
\centering
\includegraphics[width=0.975\textwidth]{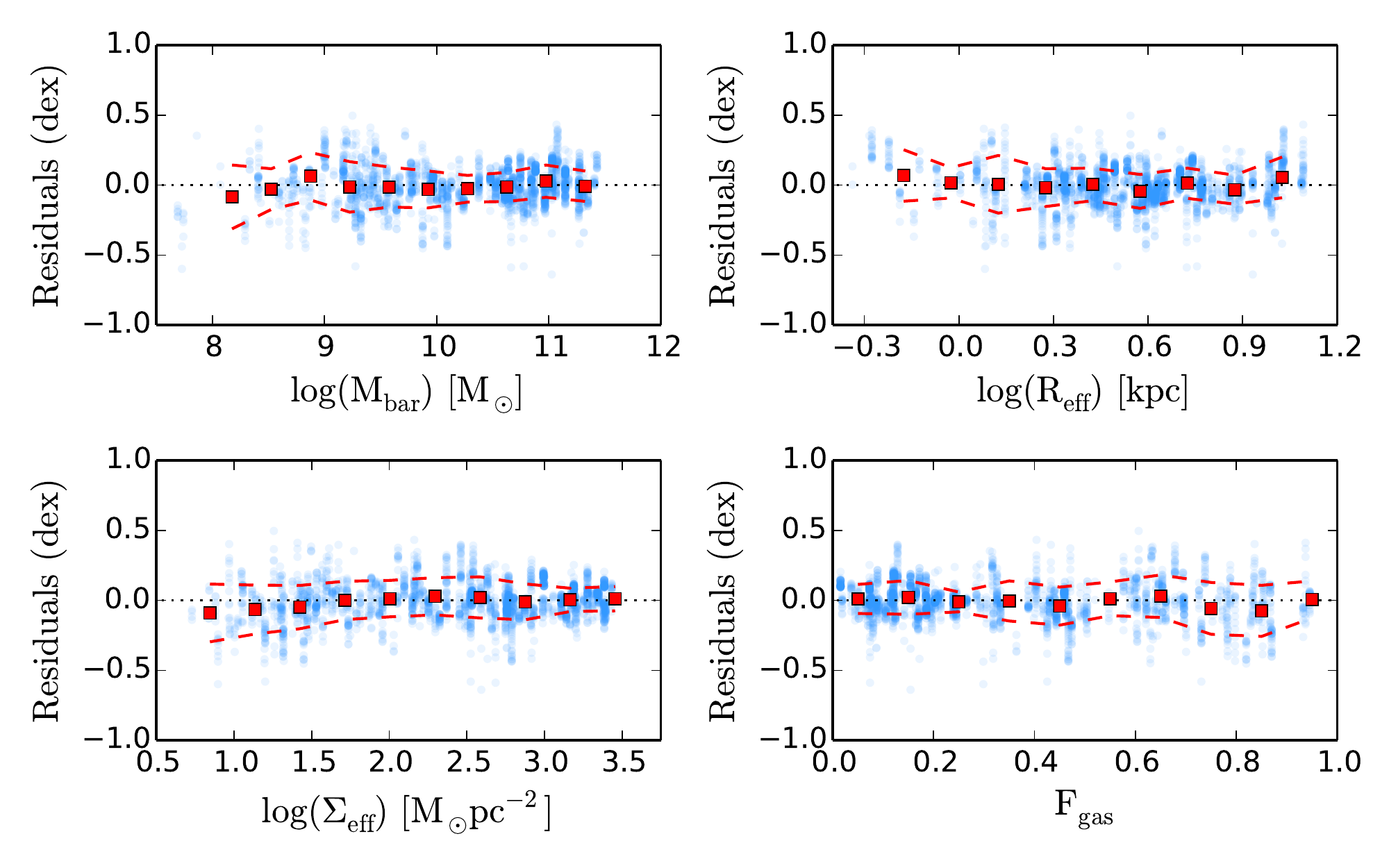}
\caption{Residuals versus baryonic mass (top left), effective radius (top right), effective surface brightness (bottom left), and global gas fraction (bottom right). Squares and dashed lines show the mean and standard deviation of binned residuals, respectively. The vertical clumps of data are due to individual objects: each galaxy contributes with several points to the radial acceleration relation.}
\label{fig:RARres2}
\end{figure*}
\subsection{Fits, Scatter, and Residuals}\label{sec:DiskFit}

We fit a generic double power-law:
\begin{equation}\label{Eq:DoublePower}
 y = \hat{y} \bigg( 1 + \dfrac{x}{\hat{x}}\bigg)^{\alpha-\beta} \bigg(\dfrac{x}{\hat{x}}\bigg)^{\beta},
\end{equation}
where $\alpha$ and $\beta$ are the asymptotic slopes for $x \gg \hat{x}$ and $x \ll \hat{x}$, respectively. We rename
\begin{equation}
 y = g_{\rm obs}, \quad x = g_{\rm bar}, \quad \hat{x} = \hat{g}_{\rm bar}, \quad \hat{y} = \hat{g}_{\rm obs}.
\end{equation}
The data are fitted using the \emph{Python} orthogonal distance regression algorithm (scipy.odr), considering errors in both variables. We do \textit{not} fit the binned data, but the individual 2693 points. The fit results are listed in Table\,\ref{tab:Fit} together with alternative choices of $\Upsilon_{\star}$ (Section\,\ref{sec:ML}) and of the dependent variable $x$ (Section\,\ref{sec:Stellar}). For our fiducial relation we find $\alpha \simeq 1$ and $\hat{g}_{\rm bar} \simeq \hat{g}_{\rm obs}$, implying that the relation is consistent with unity at high accelerations. The outer slope $\beta$ is $\sim$0.60. 

The residuals around Eq.~\ref{Eq:DoublePower}, however, are slightly asymmetric and offset from zero. The results improve by fitting the following function \citep{McGaugh2008, McGaugh2014}:
\begin{equation}\label{eq:StacyFnc}
 g_{\rm obs} = \mathcal{F}(g_{\rm bar}) = \dfrac{g_{\rm bar}}{1 - e^{-\sqrt{g_{\rm bar}/g_\dag} } },
\end{equation}
where the only free parameter is $g_\dag$. For $g_{\rm bar} \gg g_\dag$, Eq.~\ref{eq:StacyFnc} gives $g_{\rm obs} \simeq g_{\rm bar}$ in line with the values of $\alpha$, $\hat{g}_{\rm bar}$, and $\hat{g}_{\rm obs}$ found above. For $g_{\rm bar} \ll g_\dag$, Eq.~\ref{eq:StacyFnc} imposes a low-acceleration slope of 0.5. A slope of 0.5 actually provides a better fit to the low-acceleration data than 0.6 (see Fig.\,\ref{fig:RFR}). We find $g_\dag = (1.20 \pm 0.02) \times 10^{-10}$ m~s$^{-2}$. Considering a 20$\%$ uncertainty in the normalization of $\Upsilon_{[3.6]}$, the systematic error is $0.24 \times 10^{-10}$ m~s$^{-2}$.

Eq.\,\ref{eq:StacyFnc} is inspired by MOND \citep{Milgrom1983}. It is important, however, to keep data and theory well separated: Eq.\,\ref{eq:StacyFnc} is empirical and provides a convenient description of the data with a single free parameter $g_\dag$. In the specific case of MOND, the empirical constant $g_\dag$ is equivalent to the theoretical constant $a_{0}$ and $\mathcal{F}(g_{\rm bar})/g_{\rm bar}$ coincides with the interpolation function $\nu$, connecting Newtonian and Milgromian regimes. Notably, Eq.\,\ref{eq:StacyFnc} is reminiscent of the Planck Law connecting the Rayleight-Jeans and Wien regimes of electromagnetic radiation, where the Planck constant $h$ plays a similar role as $g_\dag$.

The residuals around Eq.\,\ref{eq:StacyFnc} are represented by a histogram in Fig.\,\ref{fig:RFRres} (top left). They are symmetric around zero and well fitted by a Gaussian, indicating that there are no major systematics. The Gaussian fit returns a standard deviation of 0.11 dex. This is slightly smaller than the measured rms scatter (0.13 dex) due to a few outliers. Outlying points generally come from galaxies with uncertain distances (e.g., from flow models within 15 Mpc) or poorly sampled rotation curves. Outliers are also expected if galaxies (or regions within galaxies) have $\Upsilon_{\star}$ different from our fiducial values due to, e.g., enhanced star formation activity or unusual extinction.

We constrain the observed scatter between 0.11 dex (Gaussian fit) and 0.13 dex (measured rms). These values are surprisingly small by astronomical standards. The intrinsic scatter must be even smaller since observational errors and intrinsic variations in $\Upsilon_{\star}$ are not negligible. The mean expected scatter equals to
\begin{equation}
 \dfrac{1}{N}\sum^{N}\sqrt{\delta_{\rm g_{\rm obs}}^2 + \bigg(\dfrac{\partial \mathcal{F}}{\partial g_{\rm bar}} \delta_{\rm g_{\rm bar}}\bigg)^{2}} \simeq 0.12 \, \rm{dex},
\end{equation}
where the partial derivative considers the variable slope of the relation. This explains the vast majority of the observed scatter, leaving little room for any intrinsic scatter. Clearly, the intrinsic scatter is either zero or extremely small. This is truly remarkable.

The other panels of Figure\,\ref{fig:RFRres} show the residuals versus several \textit{local} quantities: radius $R$, stellar surface density at $R$, and gas fraction $f_{\rm gas} = V_{\rm gas}^{2} / V_{\rm bar}^{2} \simeq M_{\rm gas}/M_{\rm bar}$ at $R$. The tiny residuals display no correlation with any of these quantities: the Pearson's, Spearman's, and Kendall's coefficients range between $-0.2$ and $0.1$. 

\begin{figure*}[thb]
\centering
\includegraphics[width=0.975\textwidth]{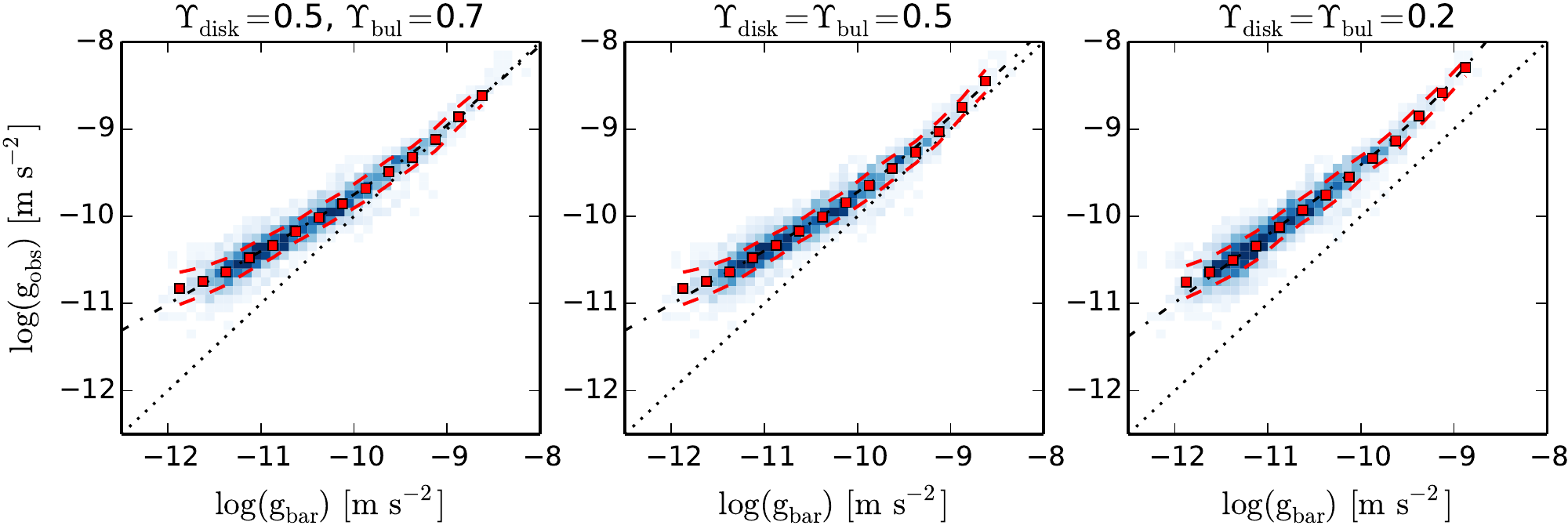}
\caption{The radial acceleration relation using different normalizations of the stellar mass-to-light ratio. Symbols are the same as in Fig.\,\ref{fig:RFR}. A relation between $g_{\rm obs}$ and $g_{\rm bar}$ persists even for low stellar mass-to-light ratio, corresponding to sub-maximal baryonic disks.}
\label{fig:ML}
\end{figure*}
\begin{figure*}[thb]
\centering
\includegraphics[width=0.975\textwidth]{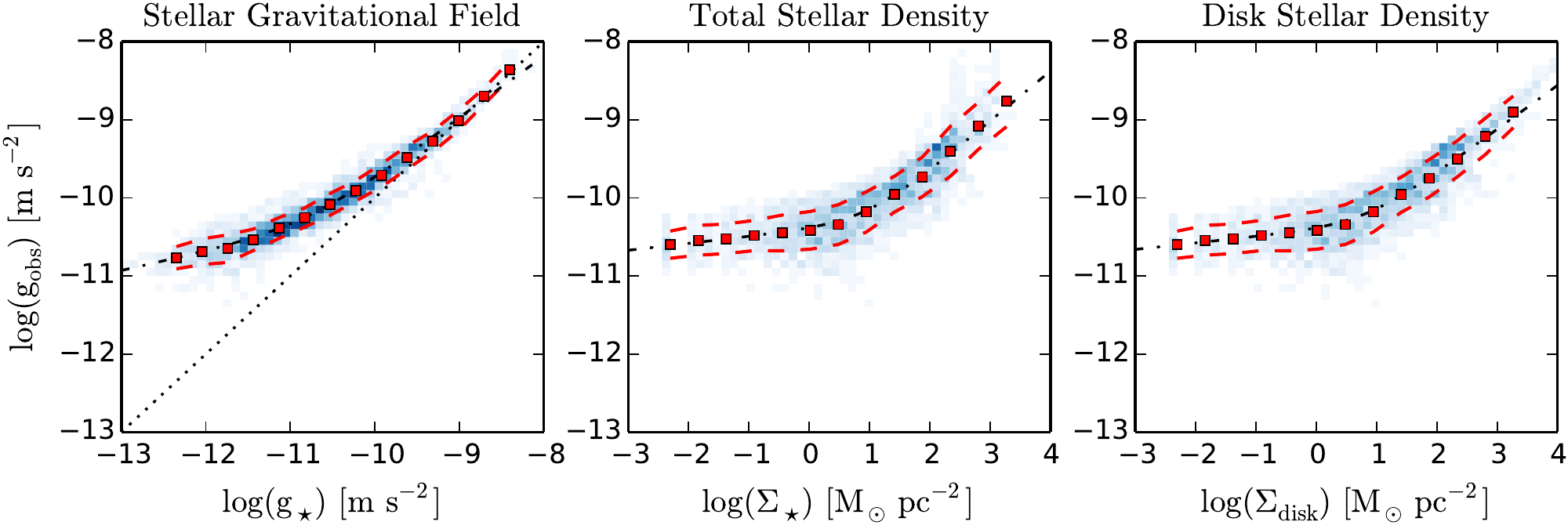}
\caption{Alternative versions of the radial acceleration relation using only photometric data. The observed acceleration at $R$ is plotted against the corresponding stellar gravitational field $g_{\star}(R)$ (left), the total stellar surface density $\Sigma_{\star}(R) = \Sigma_{\rm bul}(R) +\Sigma_{\rm disk}(R)$ (middle), and the disk stellar surface density $\Sigma_{\rm disk}(R)$ (right). Symbols are the same as in Fig.\,\ref{fig:RFR}.}
\label{fig:Alt}
\end{figure*}

Figure\,\ref{fig:RARres2} shows the residuals versus several \textit{global} quantities: baryonic mass, effective radius, effective surface brightness, and global gas fraction $F_{\rm gas} = M_{\rm gas}/M_{\rm bar}$. We do not find any statistically significant correlation with any of these quantities. Similarly, we find no correlation with flat rotation velocity, disk scale length, disk central surface brightness, and disk inclination. Strikingly, the deviations from the radial acceleration relation do not depend on any intrinsic galaxy property, either locally or globally defined.

\section{Alternative Relations for Late-Type Galaxies}

\subsection{The stellar mass-to-light ratio}\label{sec:ML}

In the previous sections, we used $\Upsilon_{\rm disk} = 0.5$ $M_{\odot}/L_{\odot}$ and $\Upsilon_{\rm bul} = 0.7$ $M_{\odot}/L_{\odot}$ at [3.6]. These values are motivated by SPS models \citep{Schombert2014a} using a \citet{Chabrier2003} IMF. They are in agreement with independent estimates from different SPS models \citep{Meidt2014, Norris2016} and resolved stellar populations in the LMC \citep{Eskew2012}. They also provide sensible gas fractions (Paper\,I) and minimize the BTFR scatter \citep{Lelli2016}. The DiskMass survey, however, reports smaller values of the $\Upsilon_{\star}$ in the $K$-band \citep{Martinsson2013, Swaters2014}, corresponding to $\sim$0.2 $M_{\odot}/L_{\odot}$ at [3.6] (see Paper\,I). Here we investigate the effects that different normalizations of $\Upsilon_{\star}$ have on the radial acceleration relation.

Figure\,\ref{fig:ML} (left) shows our fiducial relation, while Figure\,\ref{fig:ML} (middle) shows the case of $\Upsilon_{\rm disk} = \Upsilon_{\rm bul} = 0.5$ $M_{\odot}/L_{\odot}$ to isolate the role of bulges. Clearly, bulges affect only the high-acceleration portion of the relation. We fit a double power-law (Eq.\,\ref{Eq:DoublePower}) and provide fit results in Table\,\ref{tab:Fit}. The fit parameters are consistent with those from our fiducial relation, indicating that bulges play a minor role. The observed scatter increases by 0.01 dex. For $\Upsilon_{\rm bul} = 0.5$ $M_{\odot}/L_{\odot}$, the relation slighty deviates from the 1:1 line. Our fiducial value of $\Upsilon_{\rm bul} = 0.7$ $M_{\odot}/L_{\odot}$ implies that bulges are truly maximal.

\begin{figure*}[t]
\centering
\includegraphics[width=0.975\textwidth]{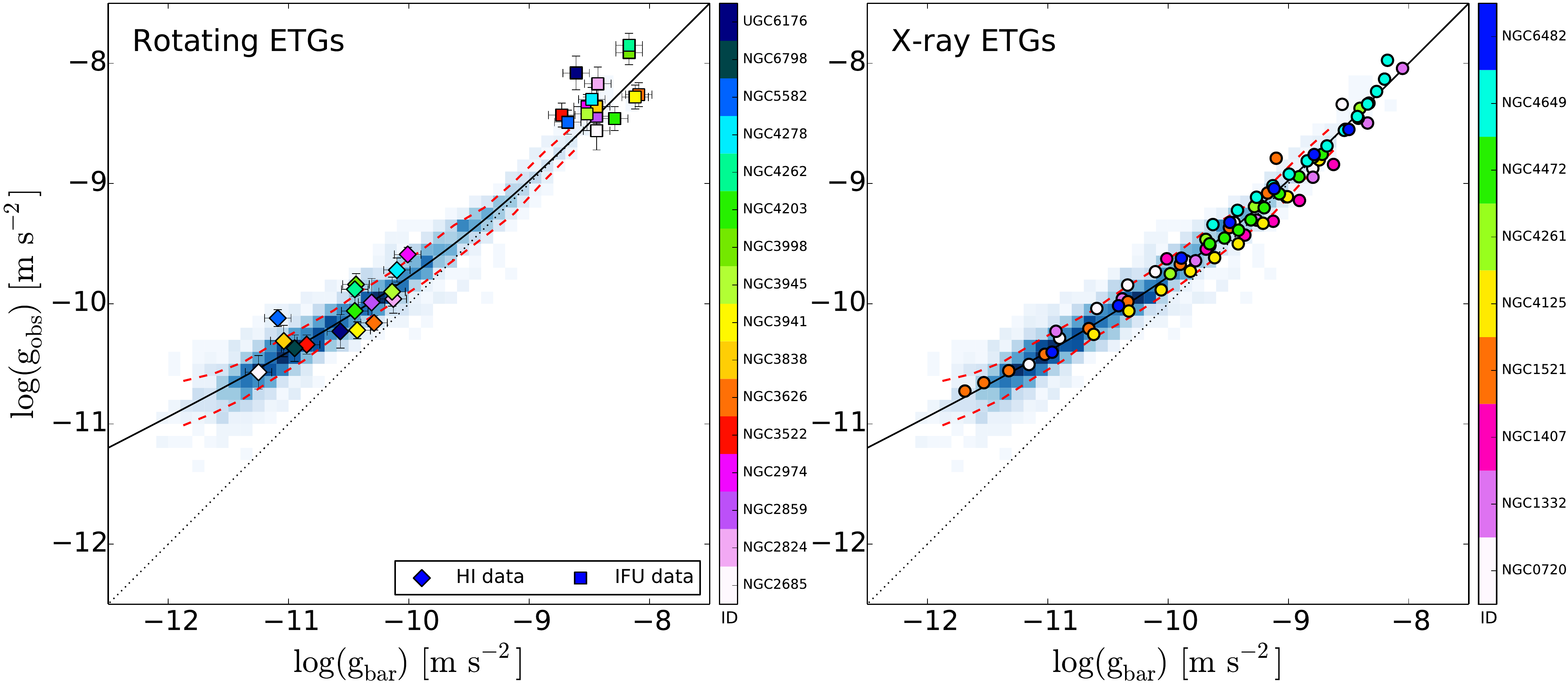}
\caption{The radial acceleration relation adding rotating ETGs from Atlas$^{\rm 3D}$ (left) and X$-$ray ETGs with accurate mass profiles (right). Each ETGs is represented with a different color as given by the legend. In the left panel, squares show measurements within 1 $R_{\rm eff}$ from JAM models of IFU data \citep{Cappellari2013}, while diamonds show measurements at large radii from \hi data \citep{denHeijer2015}. The other symbols are the same as in Fig.\,\ref{fig:RFR}. In some cases, the JAM models may over-estimate the true circular velocity (see Sect.\,\ref{sec:ETGs}).}
\label{fig:ETGs}
\end{figure*}
Figure\,\ref{fig:ML} (right) shows the radial acceleration relation using $\Upsilon_{\rm disk} = \Upsilon_{\rm bul} = 0.2$ $M_{\odot}/L_{\odot}$. For these low values of $\Upsilon_{\star}$, baryonic disks are submaximal and DM dominates everywhere (as found by the DiskMass survey). The radial acceleration relation, however, still exists: it is simply shifted in location. This happens because the shape of $V_{\rm bar}$ closely relates to the shape of $V_{\rm obs}$ as noticed early by \citet{vanAlbada1986} and \citet{Kent1987}. For $\Upsilon_{\star}=0.2$ $M_{\odot}/L_{\odot}$, the relation shows only a weak curvature since the inner and outer parts of galaxies are almost equally DM dominated. Fitting a double power-law, we find that both $\hat{g}_{\rm bar}$ and $\hat{g}_{\rm obs}$ are not well constrained and depend on the initial parameter estimates, whereas $\alpha$ and $\beta$ are slightly more stable. The values in Table\,\ref{tab:Fit} are indicative. In any case, the observed scatter significantly increases to 0.15 dex. The same happens with the BTFR \citep{Lelli2016}. The quality of both relations is negatively impacted by low $\Upsilon_{\star}$.

Interestingly, the high-acceleration slope of the relation is consistent with 1 for any choice of $\Upsilon_{\star}$ (Table\,\ref{tab:Fit}). This supports the concept of baryonic dominance in the inner parts of HSB galaxies. If HSB galaxies were strongly submaximal, they could have a slope different from unity, as seen in DM-dominated LSB galaxies. Our fiducial normalization of $\Upsilon_{\star}$ seems natural as it places the high end of the radial acceleration relation on the 1:1 line.

\subsection{Purely stellar relations}\label{sec:Stellar}

We now investigate alternative versions of the radial acceleration relation using only stellar quantities and neglecting the \hi gas. This is interesting for two reasons:
\begin{enumerate}
\item IFU surveys like CALIFA \citep{GarciaLorenzo2015} and MANGA \citep{Bundy2015} are providing large galaxy samples with spatially resolved kinematics, but they usually lack \hi data. The availability of both gas and stellar maps is one of the key advantages of SPARC. Local scaling relations between stars and dynamics, therefore, can provide a benchmark for such IFU surveys.
\item One might suspect that the radial acceleration relation is the end product of a self-regulated star-formation process, where the total gravitational field ($g_{\rm obs}$) sets the local stellar surface density. If true, one may expect that the stellar surface density correlates with $g_{\rm obs}$ better than $g_{\rm bar}$.
\end{enumerate}

In Figure \ref{fig:Alt}, we plot $g_{\rm obs}$ versus the stellar gravitational field ($g_{\star} = V_{\star}^2/R$), the total stellar surface density ($\Sigma_{\star}$), and the stellar surface density of the disk ($\Sigma_{\rm disk} = \Sigma_{\star} - \Sigma_{\rm bul}$). The tightest correlation is given by the stellar gravitational field. We recall that this quantity is obtained considering the entire run of $\Sigma_{\star}$ with radius \citep{Casertano1983}: the Newton's shell theorem does not hold for a flattened mass distribution, hence the distribution of mass \textit{outside} a given radius $R$ also affects the gravitational potential at $R$. \textit{The most fundamental relation involves the gravitational potential, not merely the local stellar density.} This cannot be trivially explained in terms of self-regulated star-formation via disk stability: it is the full run of $\Sigma_{\star}$ with $R$ (and its derivatives) to determine the quantity that best correlate with the observed acceleration at every $R$, not just the local $\Sigma_{\star}$.

The use of $g_{\star}$ instead of $g_{\rm bar}$ changes the shape of the relation: the curvature is now more acute because the data deviates more strongly from the 1:1 line below $\sim$10$^{-10}$ m~s$^{-2}$. This is the unavoidable consequence of neglecting the gas contribution. Nevertheless, the observed scatter in the $g_{\rm obs}-g_{\star}$ relation is the same as the $g_{\rm obs}-g_{\rm bar}$ relation. Hence, one can use [3.6] surface photometry to predict rotation curves with a $\sim$15$\%$ error (modulo uncertainties on distance and inclination). We perform this exercise for ETGs in the next section.

\begin{figure*}
\begin{minipage}{0.32\textwidth}
\centering
\includegraphics[width=\textwidth]{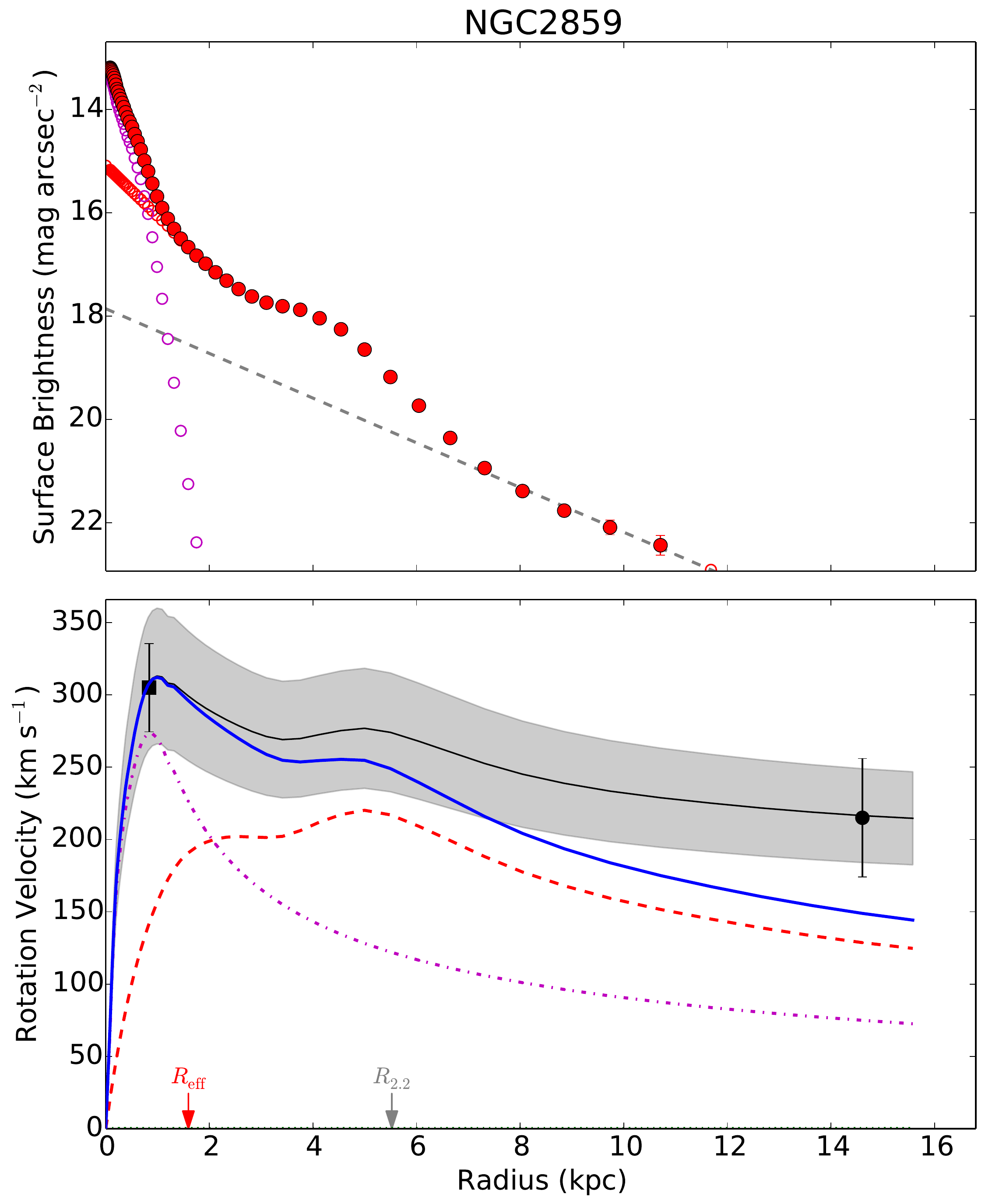}
\end{minipage}
\begin{minipage}{0.32\textwidth}
\centering
\includegraphics[width=\textwidth]{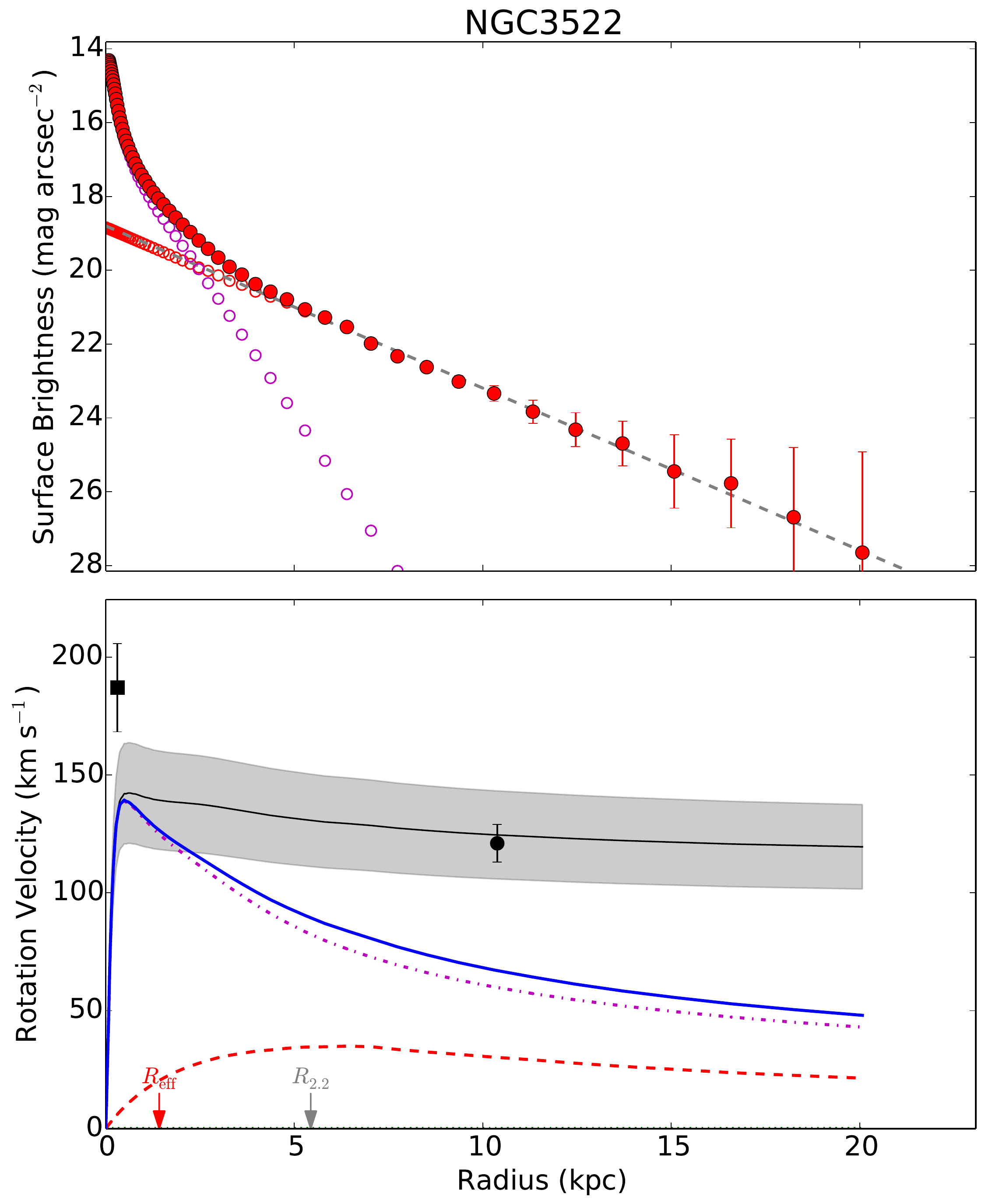}
\end{minipage}
\begin{minipage}{0.32\textwidth}
\centering
\includegraphics[width=\textwidth]{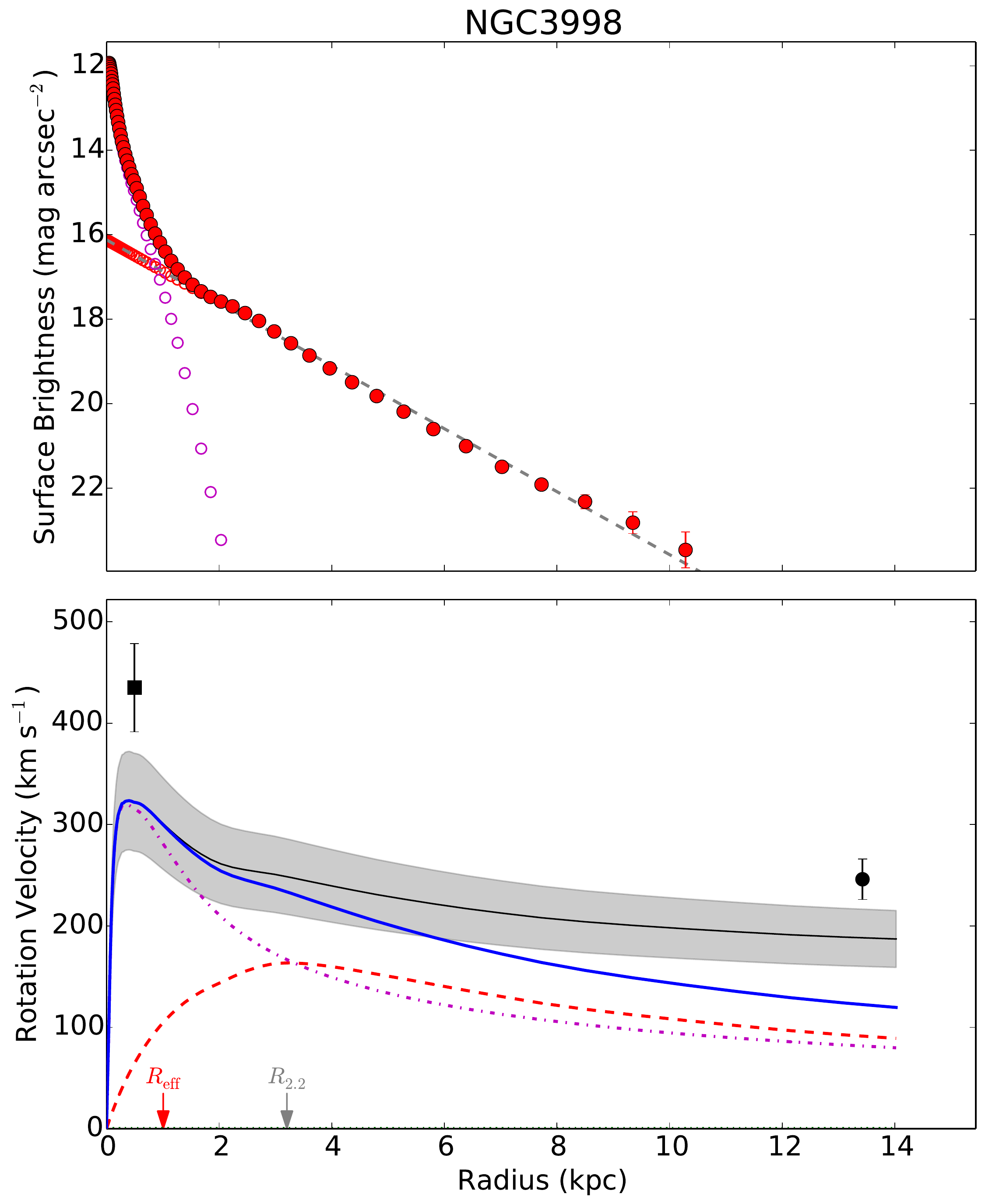}
\end{minipage}
\caption{Mass models for three rotating ETGs with extended \hi disks. \textit{Top panels}: the [3.6] surface brightness profiles (red dots) and exponential fits to the outer regions (dashed line). Open circles show extrapolated values for bulge (purple) and disk (red) components from non-parametric decompositions. \textit{Bottom panels}: velocity contributions due the bulge (purple dash-dotted line), stellar disk (red dashed line), and total baryons (blue solid line), adopting $\Upsilon_{\rm disk} = \Upsilon_{\rm bul} = 0.8$ $M_{\odot}$/$L_{\odot}$. The black line and gray band show the predicted rotation curve using the radial acceleration relation of LTGs (Eq.\,\ref{eq:StacyFnc}) and considering a scatter of 0.13 dex. The inner square shows the maximum velocity from JAM modelling of IFU data. The outer circle shows the rotation velocity from \hi data \citep{denHeijer2015}. The arrows indicate the galaxy effective radius (red) and 2.2 disk scale lengths (grey). Similar figures for all ETGs are available as a Figure Set.}
\label{fig:massmodels}
\end{figure*}
\section{Radial Acceleration Relation for Early-Type Galaxies}\label{sec:ETGs}

\subsection{General results for ETGs}

We now study the location of ETGs on the radial acceleration relation. Figure\,\ref{fig:ETGs} (left) shows rotating ETGs from Atlas$^{\rm 3D}$ with two different measurements of $g_{\rm obs}$ (Sect.\,\ref{sec:rotETGs}): one at small radii from JAM modelling of IFU data \citep{Cappellari2013} and one at large radii from \hi data \citep{denHeijer2015}. Figure\,\ref{fig:ETGs} (right) shows X$-$ray ETGs with detailed mass profiles from \emph{Chandra} and \emph{XMM} observations (Sect.\,\ref{sec:XrayETGs}): each galaxy contributes with multiple points similar to LTGs. For all ETGs, we derived $Spitzer$ [3.6] photometry and built mass models (Sect.\,\ref{sec:barETGs}): this gives estimates of $g_{\rm bar}$ that are fully independent from $g_{\rm obs}$ and comparable with those of LTGs. Our fiducial $\Upsilon_{\star}$ are listed in Table\,\ref{tab:ML}: different values would systematically shift the data in the horizontal direction without changing our overall results.

Figure \ref{fig:ETGs} shows that ETGs follow the same relation as LTGs. X$-$ray and \hi data probe a broad range of accelerations in different ETGs and nicely overlap with LTGs. The IFU data probe a narrow dynamic range at high accelerations, where we have only few measurements from bulge-dominated spirals. In general, the IFU data lie on the 1:1 relation with some scatter, reinforcing the notion that ETGs have negligible DM content in the inner parts \citep{Cappellari2016}. The same results is given by X$-$ray data at small radii, probing high acceleration regions. Some IFU data show significant deviations above the 1:1 line, but these are expected. The values of $V_{\rm max}$ from JAM models are not fully empirical and may over-estimate the true circular velocity, as we discuss below.

\citet{Davis2013} compare the predictions of JAM models with interferometric CO data using 35 ETGs with inner molecular disks. In $\sim$50$\%$ of the cases, they find good agreement. For another $\sim$20$\%$, the CO is disturbed or rotates around the polar plane of the galaxy, hence it cannot be compared with JAM models. For the remaining $\sim$30$\%$, there is poor agreement: the JAM models tend to predict systematically higher circular velocities than those observed in CO. An incidence of 30$\%$ is consistent with the deviant points in Fig.\,\ref{fig:ETGs}. We have three ETGs in common with \citet{Davis2013}. NGC~3626 (dark orange) shows good agreement between CO and JAM velocities and lie close to the 1:1 line, as expected. On the contrary, NGC~2824 (pink) and UGC~6176 (dark blue) show higher JAM velocities than observed in CO: these are among the strongest outliers in Fig.\,\ref{fig:ETGs}.

\citet{Janz2016} discuss the location of ETGs on the MDAR, which is equivalent to the radial acceleration relation after subtracting the 1:1 line. \citet{Janz2016} use JAM models of IFU data and find that ETGs follow a similar MDAR as LTGs, but with a small offset. For LTGs, they use data from \citet{McGaugh2004}, which rely on optical photometry. The apparent offset disappears using the more accurate SPARC data with [3.6] photometry. Moreover, \citet{Janz2016} use globular clusters to trace the outer gravitational potential of ETGs and find potential deviations from the MDAR at low $g_{\rm bar}$. These deviations may simply point to possible deviations from the assumed isotropy of globular cluster orbits. Both \hi and X$-$ray data show that there are no significant deviations at low accelerations: ETGs and LTGs do follow the same relation within the uncertainties.

\subsection{Predicted Rotation Curves for ETGs}

Given the uncertainties, we do not fit the radial acceleration relation including ETGs. This exercise has little value because the 2693 points of LTGs would dominate over the 28 points of rotating ETGs and 80 points of X$-$ray ETGs. Instead, we use the radial acceleration relation of LTGs (Eq.\,\ref{eq:StacyFnc}) to predict the full rotation curves of rotating ETGs. These may be tested in future studies combining CO, H$\alpha$, and \hi observations. We show three examples in Figure\,\ref{fig:massmodels}. NGC~2859 (left) exemplifies $\sim$50$\%$ of our mass models: the predicted rotation curve agrees with both IFU and \hi measurements (as expected from Fig.\,\ref{fig:ETGs}). NGC~3522 exemplifies $\sim$20$\%$ of our cases: the \hi velocity point is reproduced, but the IFU measurement is not (althought the peak radius is reproduced). This suggests that JAM models may have over-estimated the true circular velocity (as discussed above). NGC~3998 exemplifies the remaining $\sim$30$\%$: both IFU and \hi measurements are not reproduced within 1 sigma but there are systematic shifts. This may be accounted for by changing the distance, inclination, or $\Upsilon_{\star}$. 

In general, we predict that the rotation curves of ETGs should rise fast in the central regions, decline at intermediate radii, and flatten in the outer parts. The difference between peak and flat rotation velocities can be $\sim$20$\%$, e.g., 50 km~s$^{-1}$ for galaxies with $V_{\rm max}\simeq 250$ km s$^{-1}$. This is analogous to to bulge-dominated spirals \citep{Casertano1991, Noordermeer2007a}. Similar conclusions are drawn by \citet{Serra2016}.

\begin{figure*}[thb]
\centering
\includegraphics[width=0.95\textwidth]{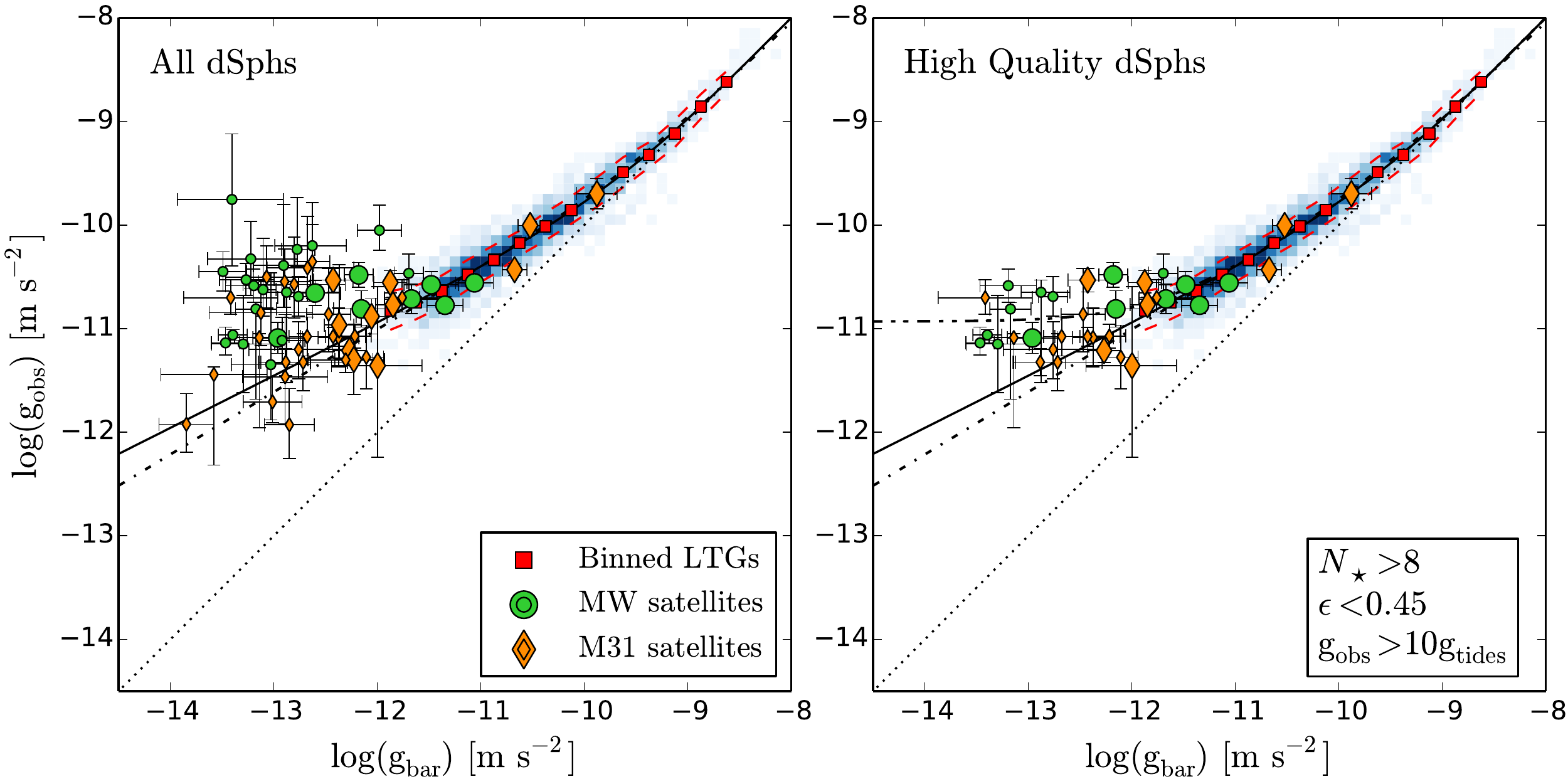}
\caption{The location of dSphs on the radial acceleration relation, using all available data (left) and data that satisfy quality criteria (right; see Sect.\,\,\ref{sec:ultrafaint}). Circles and diamonds distinguish between MW and M31 satellites. Large symbols indicate classical satellites, while small symbols show ultrafaint satellites and more isolated dSphs (like Cetus and Tucana, see e.g. \citealt{Pawlowski2014b}). The other symbols and lines are the same as in Fig.\,\ref{fig:RFR}. The dash-dotted-dotted line is a fit to Eq.\,\ref{eq:FedeFnc} using high-quality dSphs.}
\label{fig:dSphs}
\end{figure*}
\section{Radial Acceleration Relation for Dwarf Spheroidal Galaxies}\label{sec:dSphs}

\subsection{General Results for dSphs}

We now investigate the location of dSphs on the radial acceleration relation. For these objects, we have only one measurement per galaxy near the half-light radii, assuming spherical symmetry and dynamical equilibrium (see Sect.\,\ref{sec:dSphData}). These data are compiled in Appendix\,\ref{sec:Catalogue}.

In Figure\,\ref{fig:dSphs} (left), we plot all available data. Large symbols indicate the ``classical'' satellites of the MW (Carina, Draco, Fornax, Leo\,I, Leo\,II, Sculptor, Sextans, and Ursa Minor) and Andromeda (NGC\,147, NGC\,185, NGC\,205, And\,I, And\,II, And\,III, And\,V,  And\,VI, and And\,VII). Classical dSphs follow the same relation as LTGs within the errors. The brightest satellites of Andromeda adhere to the relation at $-11 \lesssim \log(g_{\rm bar}) \lesssim -10$, which are typically found in the outer parts of high-mass HSB disks or in the inner parts of low-mass LSB disks (see Fig.~\ref{fig:RFRres}, bottom). Similarly, the brightest satellites of the MW  overlap with LSB disks at $-12 \lesssim \log(g_{\rm bar}) \lesssim -11$.

In Figure\,\ref{fig:dSphs} (left), small symbols show the ``ultrafaint'' dSphs. They seem to extend the relation by a further $\sim$2 dex in $g_{\rm bar}$ but display large scatter. These objects, however, have much less accurate data than other galaxies. Photometric properties are estimated using star counts, after candidate stars are selected using color-magnitude diagrams and template isochrones. Velocity dispersions are often based on few stars and may be systematically affected by contaminants \citep{Walker2009b}, unidentified binary stars \citep{McConnachie2010}, and out-of-equilibrium kinematics \citep{McGaugh2010}. Differences of a few km~s$^{-1}$ in $\sigma_{\star}$ would significantly impact the location of ultrafaint dSphs on the plot. Hence, the shape of the radial acceleration relation at $\log(g_{\rm bar}) \lesssim -12$ remains very uncertain.

\subsection{Ultrafaint dSphs: a low-acceleration flattening?}\label{sec:ultrafaint}

As an attempt to constrain the low-acceleration shape of the relation, we apply three quality criteria:
\begin{enumerate}
 \item The velocity dispersion is measured using more than 8 stars to avoid systematics due to binary stars and velocity outliers \citep[see Appendix A.3 of][]{2013ApJ...768..172C}. This removes 8 satellites (4 of the MW and 4 of M31).
 \item The observed ellipticities are smaller than 0.45 to ensure that a spherical model (Eq.\,\ref{Eq:dSph1} and Eq.\,\ref{Eq:dSph2}) is a reasonable approximation. This removes 15 satellites (8 of the MW and 7 of M31).
 \item The tidal field of the host galaxy does not strongly affect the internal kinematics of the dSphs.
\end{enumerate}
To apply the last criterion, we estimate the tidal acceleration from the host galaxy at the half-light radii:
\begin{equation}\label{Eq:Tidal}
 g_{\rm tides} = \dfrac{G M_{\rm host}}{D_{\rm host}^{2}}\dfrac{2 r_{1/2}}{D_{\rm host}},
\end{equation}
where $D_{\rm host}$ is the distance from the host (MW or M31) and $M_{\rm host}$ is its total mass. We assume a total mass of $10^{12}$ $M_{\odot}$ for the MW and $2 \times 10^{12}$ $M_{\odot}$ for M31. These assumptions do not strongly affect our quality cut. We exclude dSphs with $g_{\rm obs}<10g_{\rm tides}$ since tides are comparable to the internal gravity. This removes four M31 satellites, reducing our dSph sample to 35 objects.

We note that the observed velocity dispersion may be inflated by non-equilibrium kinematics, hence $g_{\rm obs}$ may be over-estimated and the criterion $g_{\rm obs}<10g_{\rm tides}$ may miss some tidally affected dSphs. For example, Bootes\,I \citep{Roderick2016} and And\,XXVII \citep{2013ApJ...768..172C} show tidal features but are not excluded by our criterion. Nevertheless, these two objects are not strong outliers from the relation. Conversely, the ellipticy criterion correctly excludes some dSphs that show signs of disruption like Ursa Minor \citep{Palma2003}, Hercules \citep{Roderick2015}, and Leo~V \citep{Collins2016}.

In Figure\,\ref{fig:dSphs} (right), we enforce these quality criteria. Several outliers are removed and the scatter substantially decreases. Ultrafaint dSphs seem to trace a flattening in the relation at $\log(g_{\rm bar}) \lesssim -12$. This possible flattening is in line with the results of \citet{Strigari2008}, who found that dSphs have a constant total mass of $\sim$10$^{7}$ M$_{\odot}$ within 300 pc. Clearly, if $g_{\rm obs}\simeq$ const, the inferred mass will necessarily be constant within \textit{any} chosen radii in pc \citep[see][for a similar result]{Walker2010}

Despite the large uncertainties, it is tempting to fit the radial acceleration relation including ultrafaint dSphs. We tried both Eq.\,\ref{Eq:DoublePower} and Eq.\,\ref{eq:StacyFnc} but found unsatisfactory results because the putative low-acceleration flattening is not represented by these functional forms. Moreover, the fit is dominated by LTGs because dSphs have large errors. If we neglect the errors, we find a reasonable fit with the following function:
\begin{equation}\label{eq:FedeFnc}
 g_{\rm obs} = \dfrac{g_{\rm bar}}{1 - e^{-\sqrt{g_{\rm bar}/g_\dag} } } + \hat{g}\, e^{-\sqrt{g_{\rm bar} g_\dag/\hat{g}^2}},
\end{equation}
where the free parameters are $g_\dag$ and $\hat{g}$. This is similar to Eq.\,\ref{eq:StacyFnc} but the additional term imposes a flattening at $g_{\rm obs} = \hat{g}$ for $g_{\rm bar} < \hat{g}$. We find
\begin{equation}
\begin{split}
 g_\dag &= (1.1 \pm 0.1) \times 10^{-10} \, \mathrm{m\,s^{-2}},\\
 \hat{g} &= (9.2 \pm 0.2) \times 10^{-12} \, \mathrm{m\,s^{-2}}.
\end{split}
\end{equation}
The observed rms scatter is still small (0.14 dex). We stress that the ``acceleration floor'' expressed by $\hat{g}$ may be real or an intrinsic limitation of the current data. In a $\Lambda$CDM context, this putative low-acceleration flattening may be linked to the steep mass function of DM halos: dwarf galaxies spanning a broad range in stellar mass may form in DM halos spanning a narrow range in virial mass (as also implied by abundance matching studies). Future studies may shed new light on this issue.

\subsection{dSphs with chemo-dynamically distinct components}

The radial acceleration relation is a \textit{local} scaling law that combines data at different radii in different galaxies. In general, dSphs cannot probe this local nature because robust estimates of $g_{\rm obs}$ can only be obtained near the half-light radius, where the effects of anisotropy are small \citep[e.g.,][]{Wolf2010}. Some dSphs, however, show chemo-dynamically distinct stellar components \citep{Tolstoy2004, Battaglia2006}: one can distinguish between a metal-rich (MR) and a metal-poor (MP) component with the former more centrally concentrated and kinematically colder than the latter. These two components can be used to trace the gravitational field at different radii in the same dSph \citep{Battaglia2008, Walker2011}, thus they provide two independent points on the radial acceleration relation.

\begin{figure}[thb]
\centering
\includegraphics[width=0.47\textwidth]{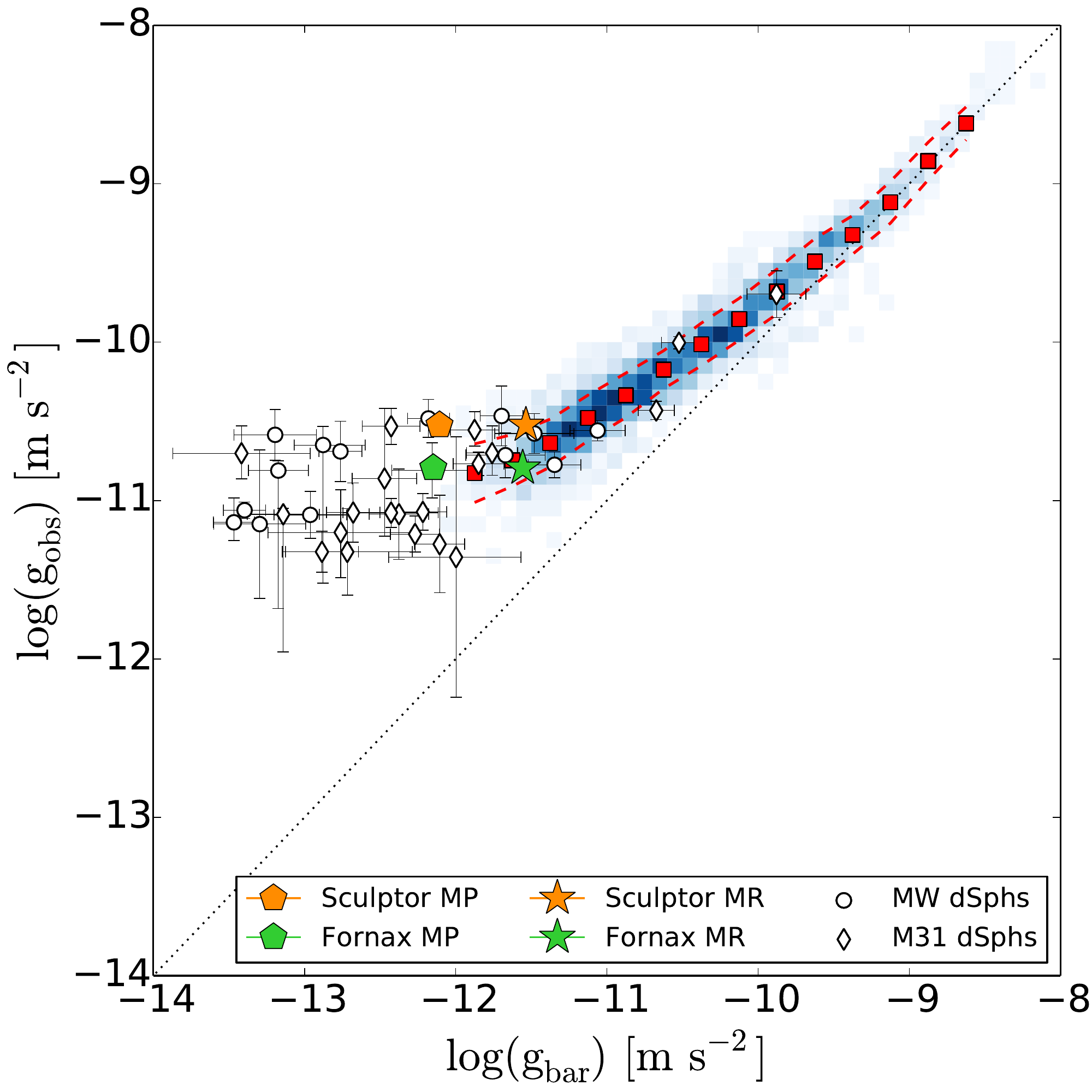}
\caption{Same as Figure\,\ref{fig:dSphs} (right) but considering the chemo-dynamically distinct components of Sculptor (red) and Fornax (green). Stars and pentagons correspond to the metal-rich and metal-poor components, respectively. These two components have different spatial distribution and kinematics, providing two measurements of $g_{\rm obs}$ at different radii.}
\label{fig:dSph2}
\end{figure}
In Figure\,\ref{fig:dSph2}, we investigate the distinct stellar components of Fornax and Sculptor using data from \citet{Walker2011}. Both components lie on the radial acceleration relation within the observed scatter, confirming its local nature. For both galaxies, the two components have approximately the same stellar mass, but the MP component has significantly larger $r_{1/2}$ leading to lower values of $g_{\rm bar}$. Even though the two components have different velocity dispersions, the value of $g_{\rm obs}$ is nearly the same. Velocity dispersions and half-light radii seem to conspire to give a constant $g_{\rm obs}$, in line with the apparent flattening of the relation at low $g_{\rm bar}$. We repeated the same exercise using different data from G. Battaglia (priv. comm.) and find only minor differences.

\begin{figure*}[thb]
\centering
\includegraphics[width=0.975\textwidth]{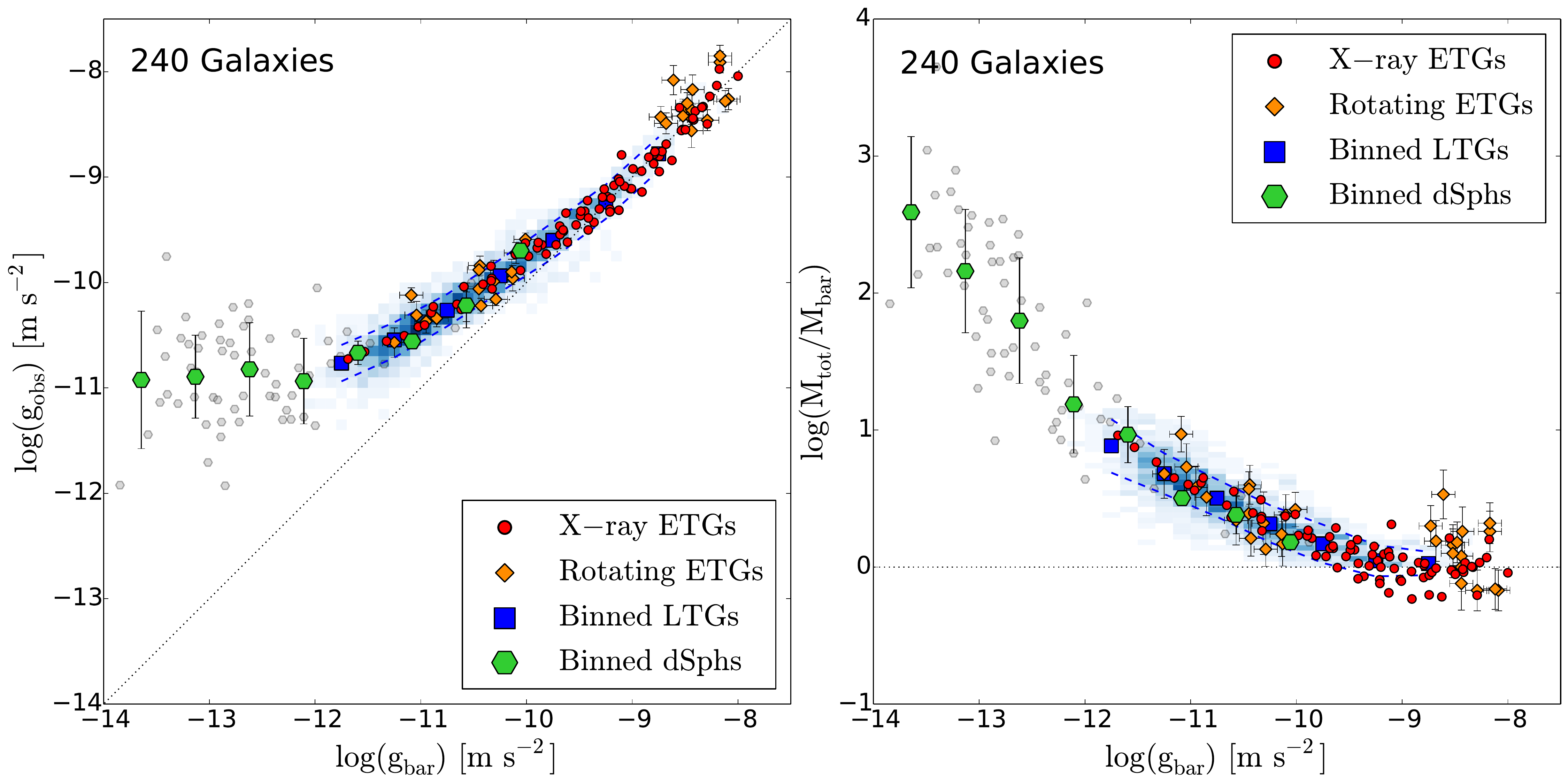}
\caption{Left: The acceleration force relation considering all galaxy types. The colorscale represents $\sim$2700 points from 153 LTGs: blue squares and dashed lines show the mean and standard deviation of binned data, respectively. Red circles and orange diamonds indicate rotating ETGs and X$-$ray ETGs, respectively. Small grey hesagons show dSphs: the large green hesagons show the mean and standard deviation of binned data. Right: The mass discrepancy $-$ acceleration relation, where the vertical axis shows $M_{\rm tot}/M_{\rm bar} \simeq g_{\rm obs}/g_{\rm bar}$. This is equivalent to subtracting the 1:1 line from the radial acceleration relation. Symbols are the same as in the left panel.}
\label{fig:Todo}
\end{figure*}
\section{Relation to Other Dynamical Laws}\label{sec:Other}

In the previous sections, we showed that LTGs, ETGs, and dSphs follow a tight radial acceleration relation: the observed acceleration correlates with that due to the distribution of baryons. This is summarized in Fig.\,\ref{fig:Todo} (left). 

To facilitate the comparison with previous works, the right panel of Figure\,\ref{fig:Todo} shows the ``mass discrepancy'' $M_{\rm tot}/M_{\rm bar} = g_{\rm obs}/g_{\rm bar}$ versus $g_{\rm bar}$. This plot is referred to as MDAR \citep{McGaugh2004, McGaugh2014}. Clearly, the mass discrepancy is not some random value but closely relates to the baryonic gravitational field. In particular, ultrafaint dSphs are heavily DM dominated ($M_{\rm tot}/M_{\rm bar}\simeq10-1000$), but their insignificant baryonic content can be used to predict $M_{\rm tot}/M_{\rm bar}$ despite the domination of DM. The observed scatter increases at high and low accelerations, but this is likely due to less precise data: the derivation of the total gravitational field for dSphs and ETGs is more complex than for LTGs due to the lack of a high-density gas disk. We also stress that the MDAR axes are no longer independent, thus it is preferable to use the radial acceleration relation.

The radial acceleration relation subsumes and generalizes several well-known dynamical properties of galaxies: (i) the BTFR \citep{McGaugh2000},(ii) the ``dichotomy'' between HSB and LSB galaxies \citep{deBlok1997, Tully1997}, (iii) the central density relation \citep{Lelli2013, Lelli2016c}, (iv) the ``baryons-halo conspiracy'' \citep{vanAlbada1986}, (v) Renzo's rule \citep{Sancisi2004}, (vi) the Faber-Jackson relation \citep{Faber1976}, and (vii) the $\sigma_{\star}-V_{\hi}$ relation \citep{Serra2016}. In the following, we discuss the interplay between these different dynamical laws.

\subsection{Baryonic Tully-Fisher Relation}

The BTFR is a consequence of the bottom-end portion of the radial acceleration relation. At large radii, we have
\begin{equation}\label{Eq:BTFR}
 g_{\rm obs}(R) \simeq \dfrac{V_{\rm f}^{2}}{R} \quad \rm{and} \quad g_{\rm bar}(R) \simeq \dfrac{G M_{\rm bar}(R)}{R^2}.
\end{equation}
The former equation is straightforward since $V_{\rm f}$ is defined as the mean value along the flat part of the rotation curve \citep[e.g.,][]{Lelli2016}. The latter equation is reasonably accurate since the monopole term typically dominates the baryonic potential beyond the bright stellar disk. The BTFR considers a single value of $V_{\rm f}$ and $M_{\rm b}$ for each galaxy. The radial acceleration relation, instead, considers each individual point along the flat part of the rotation curve and the corresponding enclosed baryonic mass. For LTGs and ETGs, the low-acceleration slope of the relation is fully consistent with 0.5, hence
\begin{equation}
 g_{\rm obs}\propto \sqrt{g_{\rm bar}} \quad \Rightarrow \quad \dfrac{V_{\rm f}^{2}}{R} \propto \dfrac{\sqrt{G M_{\rm bar}}}{R}.
\end{equation}
This eliminates the radial dependence and gives a BTFR with a slope of 4. A different bottom-end slope of the radial acceleration relation would preserve the radial dependence and imply a correlation between the BTFR residuals and some characteristic radius, contrary to the observations \citep[e.g.,][]{Lelli2016}. We stress that these results are completely empirical. Remarkably, this phenomenology was anticipated by \citet{Milgrom1983}.

\subsection{HSB-LSB Dichotomy and Central Density Relation}

HSB galaxies have steeply rising rotation curves and can be described as ``maximum disks'' in their inner parts \citep[e.g.,][]{vanAlbada1986}, whereas LSB galaxies have slowly rising rotation curves and are DM dominated at small radii \citep[e.g.,][]{deBlok1997}. \citet{Lelli2013} find that the inner slope of the rotation curve correlates with the central surface brightness, indicating that dynamical and baryonic densities are closely related. In \citet{Lelli2016c}, we estimate the central dynamical density $\Sigma_{\rm dyn}(0)$ of SPARC galaxies using a formula from \citet{Toomre1963}. We find that $\Sigma_{\rm dyn}(0)$ correlates with the central stellar density $\Sigma_{\star}(0)$ over 4 dex, leading to a central density relation \citep[see also][]{Swaters2014}.

The shape of the central density relation is similar to that of the radial acceleration relation. These two relations involve similar quantities in natural units ($G=1$), but there are major conceptual differences between them:
\begin{enumerate}
 \item The radial acceleration relation unifies points from different radii in different galaxies, whereas the central density relation relates quantities measured at $R\rightarrow0$ in every galaxy. The latter relation can be viewed as a special case of the former for $R\rightarrow0$.
 \item The Poisson's equation is applied along the ``baryonic axis'' of the radial acceleration relation ($g_{\rm bar}$), while it is used along the ``dynamical axis'' of the central density relation ($\Sigma_{\rm dyn}$) via Equation\,16 of \citet{Toomre1963}. Basically, these two relations address the same problem in reverse directions: (i) in the radial acceleration relation we start from the observed density distribution to obtain the expected dynamics $g_{\rm bar}$ and compare it with the observed dynamics $g_{\rm obs}$; (ii) in the central density relation we start from the observed dynamics to obtain the expected surface density $\Sigma_{\rm dyn}(0)$ and compare it with the observed surface density $\Sigma_{\star}(0)$.
\end{enumerate}

\subsection{Baryons-Halo Conspiracy and Renzo's Rule}

In a seminal paper, \citet{vanAlbada1986} pointed out that the rotation curves of spiral galaxies show no indication of the transition from the baryon-dominated inner regions to the DM dominated outer parts. Hence, the relative distributions of baryons and DM must ``conspire'' to keep the rotation curve flat. Similarly, the total surface density profiles of ETGs are nearly isothermal, leading to flat rotation curves and a ``bulge-halo conspiracy'' \citep[e.g.,][]{Treu2006, Humphrey2012, Cappellari2015}. The concept of ``baryons-halo conspiracy'' is embedded in the smooth shape of the radial acceleration relation, progressively deviating from the 1:1 line to the DM dominated regime at low accelerations.

Despite being remarkably flat, rotation curves can show features (bumps and wiggles), especially in their inner parts. Renzo's rule states that ``for any feature in the luminosity profile there is a corresponding feature in the rotation curve, and vice versa'' \citep{Sancisi2004}. This concept is embedded by the radial acceleration relation, linking photometry and dynamics on a radial basis. It is also generalized: the fundamental relation does not involve merely the local surface density of stars, but the local gravitational field (via the Poisson's equation) due to the entire density distribution of baryons (Sect.\,\ref{sec:Stellar}).

\subsection{Faber-Jackson and $\sigma_{\star}-V_{\hi}$ Relations}

The Faber-Jackson and $\sigma_{\star}-V_{\hi}$ relation also seems to be a consequence of the radial acceleration relation. \citet{Serra2016} considers a sample of 16 rotating ETGs and find that $\sigma_{\star}$ linearly correlates with $V_{\hi}$, where $V_{\hi}\simeq V_{\rm flat}$ \citep{denHeijer2015}. They report
\begin{equation}
 V_{\hi} = 1.33 \, \sigma_{\star}
\end{equation}
with an observed scatter of only 0.05 dex. The extremely small scatter suggests that the $\sigma_{\star}-V_{\hi}$ relation is most fundamental for ETGs \citep[see also][]{Pizzella2005, Courteau2007}. \citet{Serra2016} discuss that the $\sigma_{\star}-V_{\hi}$ relation implies a close link between the inner baryon-dominated regions (probed by $\sigma_{\star}$ or $V_{\rm max}$) and the outer DM-dominated parts (probed by $V_{\hi}$). This link is made explicit in Figure\,\ref{fig:ETGs}: ETGs follow the same radial acceleration relation as LTGs, combining inner and outer parts of galaxies into a smooth single relation.

\citet{denHeijer2015} show that rotating ETGs follow the same BTFR as LTGs: 
\begin{equation}
M_{\rm bar} = N \, V_{\hi}^4.
\end{equation}
Hence, a baryonic Faber-Jackson relation follows:
\begin{equation}
M_{\rm bar} = N \, (1.33)^4 \, \sigma_{\star}^4. 
\end{equation}
The radial acceleration relation appears to be the most fundamental scaling law of galaxies, encompassing previously known dynamical properties. It links the central density relation at $R=0$ and the BTFR at large radii. The smooth connection between the baryon dominated and DM dominated regimes generalizes concepts like disk-halo conspiracy and Renzo's rule for LTGs, as well as the Faber-Jackson and $\sigma_{\star}-V_{\hi}$ relations for ETGs. The radial acceleration relation is tantamount to a Natural Law: a sort of Kepler's law for galaxies.

\section{Implications}\label{sec:Impl}

\subsection{Implications for the DM distribution}

In this work we have not considered any specific model for the DM halo. Empirically, there is no reason to do so. For LTGs and ETGs, the detailed DM distribution directly follows from the radial acceleration relation. The gravitational field due to DM can be written entirely in terms of the baryonic field (see also \citealt{McGaugh2004}):
\begin{equation}\label{Eq:DMforce}
 g_{\rm DM} = g_{\rm tot} - g_{\rm bar} = \mathcal{F}(g_{\rm bar}) - g_{\rm bar},
\end{equation}
where $\mathcal{F}(g_{\rm bar})$ is given by either Eq.\,\ref{Eq:DoublePower} or Eq.\,\ref{eq:StacyFnc}. Note that Eq.\,\ref{eq:FedeFnc} is unrealistic in this context because it would imply that rotation curves start to rise as $\sqrt{R}$ beyond $\hat{g} \simeq 10^{-11}$ m~s$^{-2}$. Eq.\,\ref{eq:FedeFnc} is just a convenient function to include ultrafaint dSphs in the radial acceleration relation. For a spherical DM halo, the enclosed mass is
\begin{equation}\label{Eq:DMmass}
 M_{\rm DM}(< R) = \dfrac{R^2}{G} \bigg[\mathcal{F}(g_{\rm bar})- g_{\rm bar}\bigg].
\end{equation}
The observed baryonic distribution specifies the DM distribution. There is no need to fit arbitrary halo models.

\subsection{Implications for Galaxy Formation Models}

\subsubsection{General implications and conceptual issues}

In a $\Lambda$CDM cosmology, the process of galaxy formation is highly stochastic. DM halos grow by hierarchical merging and accrete gas via hot or cold modes, depending on their mass and redshift \citep{Dekel2006}. Gas is converted into stars via starbursts or self-regulated processes, leading to diverse star-formation histories. Supernova explosions, stellar winds, and active galactic nuclei can inject energy into the ISM and potentially drive gas outflows, redistributing both mass and angular momentum \citep{Governato2010, Madau2014, DiCintio2014}. Despite the complex and diverse nature of these processes, galaxies follow tight scaling laws. Regularity must somehow emerge from stochasticity. This issue is already puzzling in the context of global scaling laws, like the Tully-Fisher and Faber-Jackson relations, but is exacerbated in the context of the radial acceleration relation due to its local nature.

\begin{figure}[thb]
\centering
\includegraphics[width=0.48\textwidth]{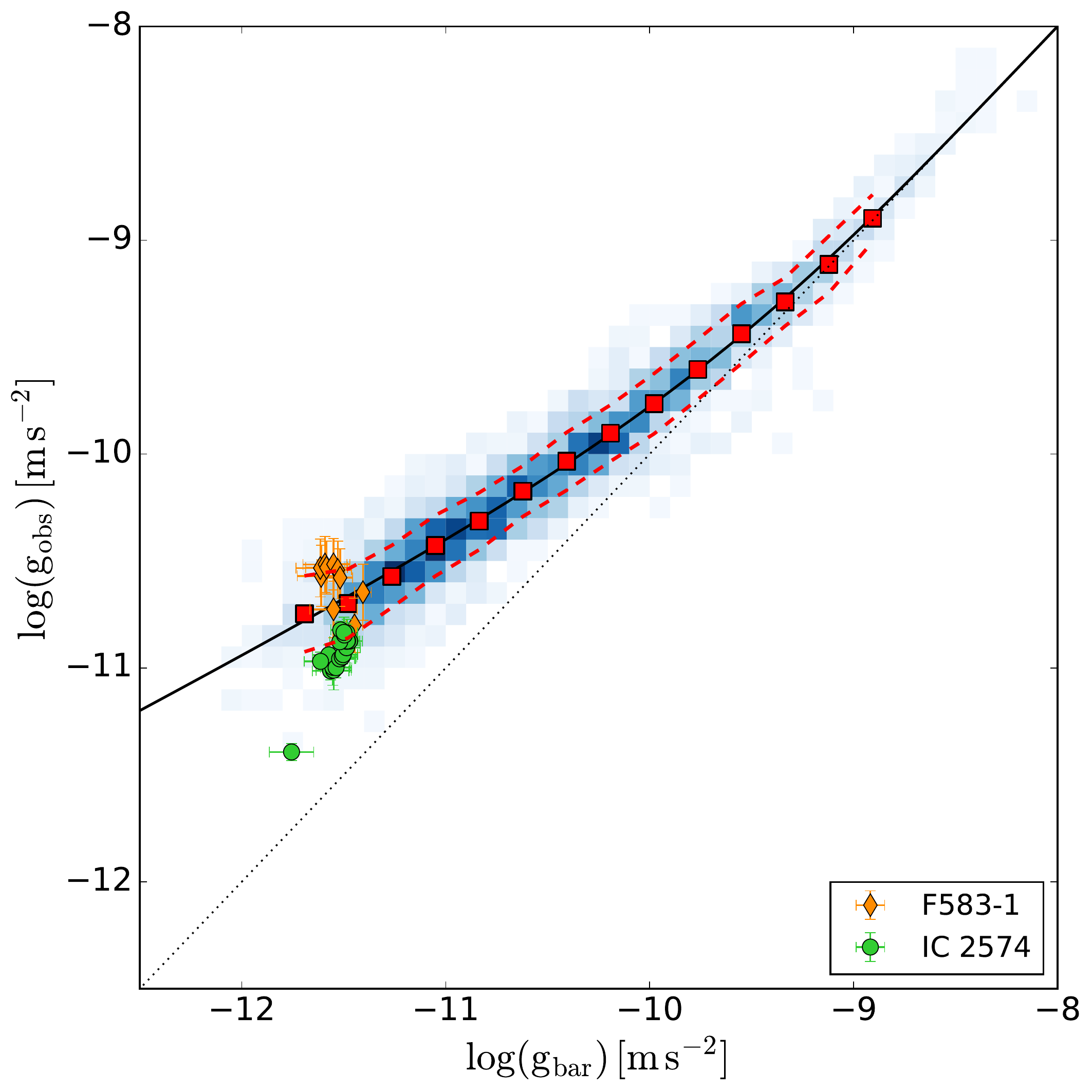}
\caption{Same as Figure\,\ref{fig:RFR} (top left) but we highlight F583-1 and IC\,2574. These two LSB galaxies are identified by \citet{Oman2015} as challenging cases for $\Lambda$CDM, yet they fall on the radial acceleration relation within the observed scatter. The shifts with respect to the main relation may be due to variations in $\Upsilon_{\star}$.}
\label{fig:Oman}
\end{figure}

The $\Lambda$CDM model is known to face severe problems with both the rotation curves of disk galaxies \citep[e.g.,][]{Moore1999, deBlok2001, McGaugh2007, Kuzio2009} and the dynamics of dSphs \citep[e.g.,][]{Boylan-Kolchin2011, Walker2011, Pawlowski2014}. Recently, \citet{Oman2015} pointed out the ``unexpected diversity of dwarf galaxy rotation curves''. This statement is correct from a pure $\Lambda$CDM perspective since little variation is expected in the DM distribution at a given mass. However, it misses a key observational fact: the diversity of rotation curve shapes is fully accounted by the diversity in baryonic mass distributions \citep{Sancisi2004, Swaters2012, Lelli2013, Karukes2017}. For example, IC\,2574 and F583-1 are identified by \citet{Oman2015} as problematic cases for $\Lambda$CDM (see their Fig.\,5). Still, these galaxies fall on the observed radial acceleration relation (Fig.\,\ref{fig:Oman}). From an empirical perspective, their slowly rising rotation curves are fully expected due to their low surface brightnesses \citep{Lelli2013}. The real problem is not ``diversity'', but the remarkable regularity in the baryon-DM coupling. Similarly, the ``cusp-core'' and ``too-big-to-fail'' problems are merely symptoms of a broader issue for $\Lambda$CDM: rotation curves can be predicted from the baryonic distribution, even for galaxies that appear to be entirely DM dominated!

A satisfactory model of galaxy formation should explain four conceptual issues: (i) the physical origin of the acceleration scale $g_{\dag}$, (ii) the physical origin of the low-acceleration slope (consistent with 0.5), (iii) the intrinsic tightness of the relation, and (iv) the lack of correlations between residuals and other galaxy properties.

\subsubsection{Comparison with hydrodynamical simulations}

\citet{Wu2015} compared the data of \citet{McGaugh2004} with cosmological simulations of galaxy formation \citep{Agertz2011, Guedes2011, Aumer2013, Marinacci2014}, finding poor agreement. On the other hand, \citet{Santos2016} find reasonable agreement between the data of \citet{McGaugh2014} and 22 model galaxies from zoom-in simulations of small volumes. This is encouraging, but more simulated galaxies are needed to quantify the theoretical scatter around these relations and have a proper comparison with observations. For example, \citet{Keller2016} find good agreement between the SPARC relation and 18 simulated galaxies, but their model galaxies span only a factor of 15 in mass covering less than 0.05$\%$ of the SPARC mass range. To conclude that ``$\Lambda$CDM is fully consistent with the SPARC acceleration law'', the simulated galaxies should at least span the same ranges in luminosity, surface brightness, size, and gas fraction as the data.

Recently, \citet{Ludlow2016} analysed simulated galaxies from the EAGLE and APOSTOLE projects. This work significantly increases the statistics and mass range with respect to \citet{Santos2016} and \citet{Keller2016}, but present several problems:
\begin{enumerate}
 \item \citet{Ludlow2016} fit the simulated data using our Eq.\,\ref{eq:StacyFnc} but find $g_{\dag}=3.00$ instead of $g_{\dag} = 1.20 \pm 0.02\,\mathrm{(rnd)} \pm 0.24 \,\mathrm{(sys)}$. The discrepancy is 90$\sigma$ (rnd) and 7.5$\sigma$ (sys). This indicates that simulations predict too much DM in galaxies, which is a persistent problem for $\Lambda$CDM \citep[e.g.,][]{McGaugh2007, Kuzio2009}. Indeed, the same simulations were used by the same group to reach the opposite conclusion: many real galaxies show a putative ``inner mass deficit'' with respect to $\Lambda$CDM expectations \citep{Oman2015}.
 \item \citet{Ludlow2016} compare the \textit{theoretical} scatter from the numerical simulations with the \textit{observed} scatter. This is not appropriate because the observed scatter is largely driven by observational errors (see Sect.\,\ref{sec:DiskFit} and \citealt{McGaugh2016}). One should compare to the \textit{intrinsic} scatter, which is either zero or extremely small ($\lesssim$0.05 dex). The theoretical scatter from \citet{Ludlow2016}, though small (0.09 dex), is still too large compared to the observations.
 \item \citet{Ludlow2016} compute $g_{\rm obs}$ and $g_{\rm bar}$ assuming spherical symmetry instead of estimating the gravitational potential in the disk mid-plane. The difference between spherical and disk geometry is not terribly large but significant. This introduces systematics that are hard to address.
 \item According to their Figure 3 (left panel), there is a systematic off-set between high-mass galaxies from EAGLE and low-mass galaxies from APOSTOLE. We do not observe any off-set between high and low mass galaxies (see Fig.\,\ref{fig:RFR} and Fig.\,\ref{fig:RARres2}).
\end{enumerate}
In summary, several key properties of the radial acceleration relation are not reproduced by the current generation of cosmological simulations. The claim that this relation is ``a natural outcome of galaxy formation in CDM halos'' has yet to be demonstrated.

\subsubsection{Comparison with semi-empirical models}

\citet{DiCintio2016} investigated the BTFR and MDAR using a simple, analytic, semi-empirical model. They assign disk galaxies to DM halos using abundance matching prescriptions and specify the distribution of gas and stars using observed scaling relations. This is the most optimistic $\Lambda$CDM model imaginable: baryonic physics is assumed to work just right to form realistic galaxies, while the properties of DM halos are taken from simulations (with or without the effects of baryons). This model can reproduce the overall shape of the MDAR, hence the radial acceleration relation. \textit{This is not trivial: the DM fractions from abundance matching depend on the relation between the theoretical halo mass function and the observed stellar mass function. There is no guarantee that they should reproduce the mass discrepencies observed in real galaxies.}

While promising, the model of \citet{DiCintio2016} predicts significant intrinsic scatter around the MDAR and correlations between residuals and radius (see their Figure 2). In $\Lambda$CDM the intrinsic scatter is driven by scatter in the baryonic-to-halo mass ratio, mass-concentration relation of DM halos, mass-size relation of stellar disks, etc. These sources of scatter are unavoidable in any hierarchical DM model of galaxy formation. Similar results are found by \citet{Desmond2017} using a similar semi-empirical approach and a sophisticated statistical framework: even models with zero scatter on abundance matching prescriptions predict too large scatter!  

The radial acceleration relation cannot be an accident of faulty data. If there were systematic errors (due to, e.g., beam smearing), even a tight intrinsic relation would be washed out. In contrast, the only way to increase the intrinsic scatter is if we have \textit{overestimated} the uncertainties. We are confident that our errors are not overestimated. We assume relatively small variations in $\Upsilon_{\star}$ (0.11 dex) and neglect some sources of error like that in the vertical density distribution of disks and bulges. There is simply little room for intrinsic scatter.

\subsection{Implications for Alternative Theories}

The tightness of the radial acceleration relation and the lack of residual correlations may suggest the need of a revision of the standard DM paradigm. We envisage two general scenarios: (I) we need new fundamental laws of physics rather than DM, or (II) we need new physics in the dark sector leading to a baryon-DM coupling. 

\subsubsection{New laws of physics rather than dark matter?}

Scenario I includes MOND \citep{Milgrom1983}, modified gravity \citep{Moffat2016}, entropic gravity \citep{Verlinde2016}, conformal gravity \citep{OBrien2012}, and fifth forces \citep{Burrage2016}. Strikingly, \citet{Milgrom1983} predicted the existence of the radial acceleration relation 33 years ago when only a few rotation curves were available \citep{Rubin1978, Bosma1978} and no detailed mass models were built. MOND can be viewed either as modified gravity (MG) by changing the Poisson's equation \citep{Bekenstein1984, Milgrom2010} or modified inertia (MI) by changing the Second Law of Newton \citep{Milgrom1994, Milgrom2006}. Several relativistic extensions of MOND have been proposed \citep[see][]{Famaey2012}, but it is unclear whether they can reproduce the cosmic microwave background and the formation of cosmic structures \citep{McGaugh2015b}.

The basic tenet of MOND is that dynamics become \textit{scale-invariant} below a critical acceleration scale $a_0$ \citep{Milgrom2009}: the equations of motion remain unchanged under the transformation $(\vec{r}, t) \rightarrow (\lambda \vec{r}, \lambda t)$. Scale-invariance specifies the bottom-end behaviour of the radial acceleration relation: $g_{\rm obs} = \sqrt{\alpha a_{0} g_{\rm bar}}$, where $\alpha=1$ for test particles on circular orbits far from a given mass distribution \citep{Milgrom2014}. In MG theories, $\alpha =1$ for all orbits in all spherical systems, but it could vary from galaxy to galaxy (or even within the same galaxy) in nonspherical systems like disk galaxies. In MI theories, $\alpha=1$ for circular orbits in dynamically-cold systems like \hi disks, but it may differ from unity if there are chaotic orbits or significant pressure support, like in the stellar components of ETGs and dSphs.

For LTGs, ETGs, and classic dSphs, we find that the relation is consistent with a low-acceleration slope of 0.5 and $\alpha=1$ for every galaxy, in agreement with general MOND predictions. For ultrafaint dSphs, the outer slope seems to be shallower than 0.5 (Figure\,\ref{fig:dSphs}). This result must be taken with caution because (i) the data of ultrafaint dSphs are much more uncertain than those of other galaxies, (ii) we cannot robustly estimate $g_{\rm obs}$ at different radii (as in LTGs and ETGs) and test possible variations of $\alpha$ from galaxy to galaxy, and (iii) in MOND we expect many dSphs to be affected by the external field effect (EFE), as we discuss below. In any case, Figure\,\ref{fig:dSphs} suggests some sort of organizing principle.

We note that the intrinsic scatter around the radial acceleration relation could distinguish between MI and MG theories. In MI theories, the relation $g_{\rm obs} = \nu(g_{\rm bar}/a_{0}) g_{\rm bar}$ holds exactly for circular orbits  \citep{Milgrom1994}, hence the radial acceleration relation should have zero intrinsic scatter. In MG theories, the above relation is valid only for spherically symmetric systems \citep{Bekenstein1984}. For flattened systems like LTGs, the predicted $g_{\rm obs}$ can show significant differences with respect to the algebraic relation \citep{Brada1995}, potentially causing some small intrinsic scatter in the radial acceleration relation. The current data cannot distinguish between tiny or null intrinsic scatter.

\subsubsection{New physics in the dark sector?}

Scenario II includes theories like dark fluids \citep{Zhao2010, Khoury2015} and dipolar DM particles subjected to gravitational polarization \citep{Blanchet2008, Blanchet2009}. By construction, these theories reconcile the successes of $\Lambda$CDM on cosmological scales with those of MOND on galaxy scales, hence they can explain the radial acceleration relation for LTGs and ETGs. The situation is more complex for dSphs because they are not isolated and the internal gravitational field can be comparable to the gravitational field from the host galaxy ($g_{\rm host}$). In pure MOND theories (scenario I), this leads to the EFE: the strong equivalence principle is violated and the internal dynamics of a system can be affected by an external field \citep{Bekenstein1984}. This is not necessarily the case in hybrid theories (scenario II), hence dSphs may help distinguish between fundamental dynamics (I) or new dark sector physics (II).

\begin{figure}[thb]
\centering
\includegraphics[width=0.48\textwidth]{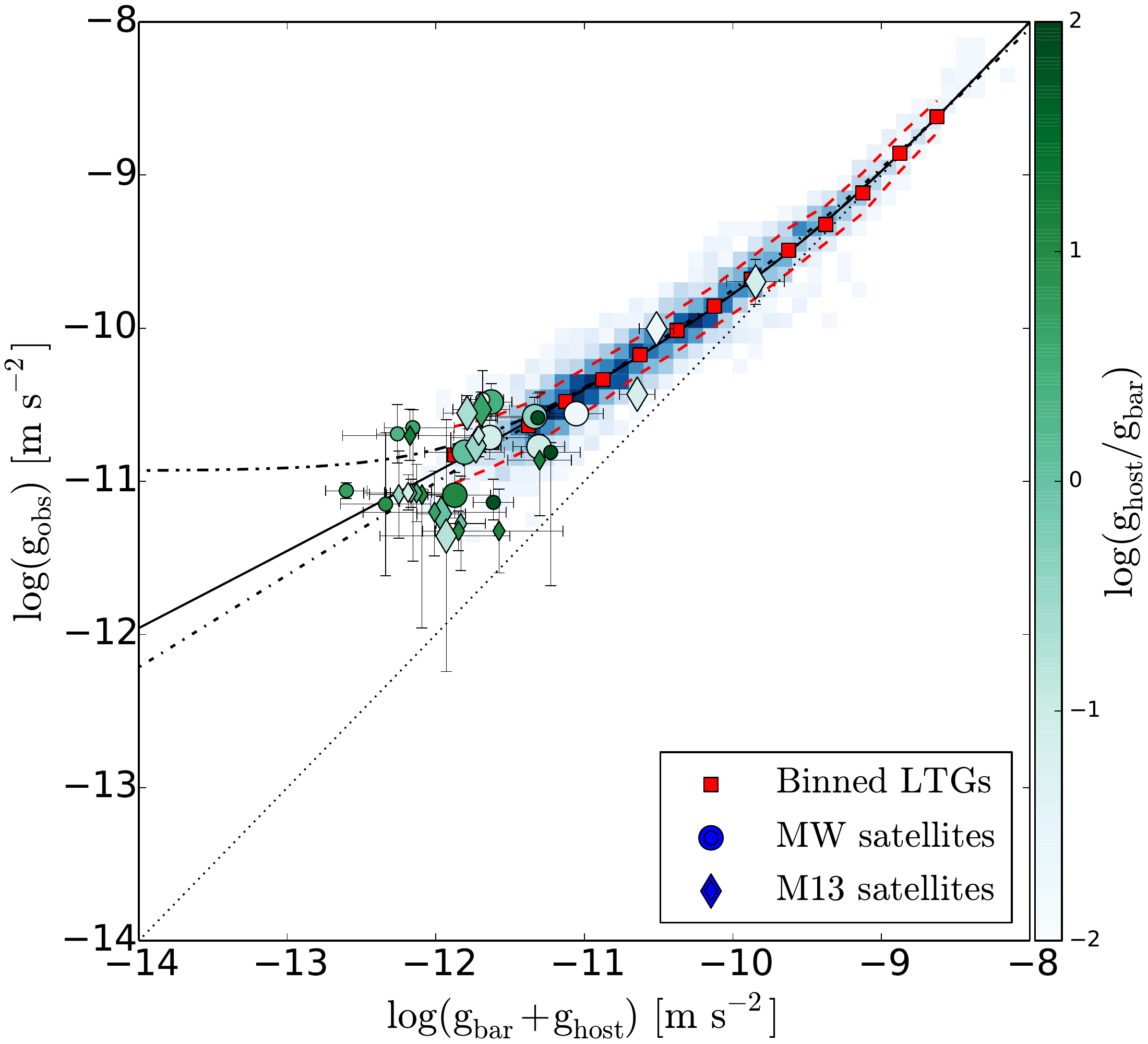}
\caption{The radial acceleration relation replacing $g_{\rm bar}$ with $g_{\rm bar} + g_{\rm host}$. $g_{\rm host}$ is the baryonic gravitational field from the host galaxy (either M31 or MW). dSphs are color-coded by the ratio $g_{\rm host}/g_{\rm bar}$ which quantifies the entity of the horizontal shift. The other symbols are the same as in Fig.\,\ref{fig:dSphs}.}
\label{fig:EFE}
\end{figure}

Inspired by the EFE, we plot the radial acceleration relation replacing $g_{\rm bar}$ with $g_{\rm bar}+g_{\rm host}$ (Figure \ref{fig:EFE}). For LTGs, we assume $g_{\rm host} = 0$ as these are relatively isolated systems. For dSphs, we estimate $g_{\rm host}$ as
\begin{equation}
g_{\rm host} = \dfrac{G M_{\rm host}}{D_{\rm host}^2},
\end{equation}
where $D_{\rm host}$ has the same meaning as in Sect.\,\ref{sec:dSphs} but $M_{\rm host}$ is now the baryonic mass of the host. Strikingly, ultrafaint dSphs now lie on the same relation as more luminous galaxies. Note that $g_{\rm host}$ depends on the distance from the host and can vary by $\sim$2 dex from galaxy to galaxy. It is surprising that this variable factor shifts dSphs roughly on top of the relation for LTGs. 

We stress, however, that this is \textit{not} a standard MOND implementation of the EFE. In the quasi-linear formulation of MOND \citep{Milgrom2010}, the EFE can be approximated by Eq.\,60 of \citet{Famaey2012}:
\begin{equation}\label{eq:EFE}
 g_{\rm obs} = \big(g_{\rm bar} + g_{\rm host}\big) \, \nu\bigg(\dfrac{g_{\rm bar} + g_{\rm host}}{a_0}\bigg) - g_{\rm host} \,\nu\bigg(\dfrac{g_{\rm host}}{a_0}\bigg),
\end{equation}
where $\nu$ is the interpolation function (equivalent to the shape of the radial acceleration relation). Basically, Figure\,\ref{fig:EFE} neglects the negative term in Eq.\,\ref{eq:EFE}, driving non-linearity in $g_{\rm tot} = g_{\rm bar} + g_{\rm host}$. This term is important because it ensures the required Newtonian limit: $g_{\rm obs} = g_{\rm bar}$ for $g_{\rm host} \gg a_0 \gg g_{\rm bar}$. Neglecting this term, we would have $g_{\rm obs} = g_{\rm bar} + g_{\rm host}$ when either $g_{\rm host} \gg a_0$ or $g_{\rm bar} \gg a_0$, leading to some strange results. For example, the internal dynamics of globular clusters would depend on their location within the MW. The location of ultrafaint dSphs in Figure\,\ref{fig:EFE} may be a coincidence or perhaps hint at some deeper physical meaning.

\section{Summary \& Conclusions}\label{sec:Conc}

We study \textit{local} scaling relations between baryons and dynamics in galaxies. Our sample includes 240 galaxies with spatially resolved kinematic data, spanning $\sim$9 dex in baryonic mass and covering all morphological types. We consider (i) 153 LTGs (spirals and irregulars) from the SPARC database, (i) 25 ETGs (ellipticals and lenticulars) with stellar and \hi kinematics from Atlas$^{\rm 3D}$ or X$-$ray observations from \emph{Chandra} and \emph{XMM}, and (iii) 62 dSphs in the Local Group with individual-star spectroscopy. Our results can be summarized as follows:
\begin{enumerate}
\item For LTGs the observed acceleration (from the rotation curve) correlates with the baryonic gravitational field (from the distributions of gas and stars) over $\sim$4 dex. This relation combines $\sim$2700 datapoints at different radii in different galaxies. The outer parts of high-mass HSB galaxies smoothly overlap with the inner parts of low-mass LSB ones.
\item The observed scatter is very small ($\lesssim$0.13 dex) and largely driven by observational uncertainties: the radial acceleration relation has little (if any) intrinsic scatter. The tiny residuals show no correlation with either local or global galaxy properties. There is no hint of a second parameter.
\item The radial acceleration relation holds for any reasonable choice of $\Upsilon_{\star}$. The high-end slope is always consistent with 1, suggesting that HSB galaxies are baryon dominated in their inner parts.
\item The stellar surface density correlates with $g_{\rm obs}$ less tightly than $g_{\rm bar}$ or $g_{\star}$. The fundamental relation involves the baryonic gravitational field via the Poisson's equation, considering the entire surface density distribution, not merely the local surface density at $R$.  
\item ETGs follow the same radial acceleration relation as LTGs. We predict the full rotation curves of rotating ETGs using the radial acceleration relation of LTGs and the observed [3.6] luminosity profiles.   
 \item The most luminous dSphs overlap with LTGs and ETGs on the radial acceleration relation, whereas ultrafaint dSphs seem to extend the relation $\sim$2 dex lower in $g_{\rm bar}$. If the data are trimmed with basic quality criteria, ultrafaint dSphs may possibly trace a low-acceleration flattening in the relation. If confirmed, this would explain the constant value of the total mass found by \citet{Strigari2008}.
 \end{enumerate}
 
The radial acceleration relation describes the \textit{local} link between baryons and dynamics in galaxies, encompassing and generalizing several well-known galaxy scaling laws. This is tantamount to a Natural Law: a sort of Kepler law for galactic systems. A tight coupling between baryons and DM is difficult to understand within the standard $\Lambda$CDM cosmology. Our results may point to the need for a revision of the current DM paradigm.

\acknowledgments

FL thanks G. Battaglia, F. Fraternali, G. Pezzulli, R. Sancisi, and M. Verheijen for many inspiring discussions over the years. Support for this work was provided by NASA through Hubble Fellowship grant \#HST-HF2-51379.001-A awarded by the Space Telescope Science Institute, which is operated by the Association of Universities for Research in Astronomy, Inc., for NASA, under contract NAS5-26555. This publication was made possible through the support of a grant from the John Templeton  Foundation. The opinions expressed in this publication are those of the authors and do not necessarily reflect the views of the John Templeton Foundation.

\appendix

\begin{table*}
\begin{center}
\caption{Sample of early-type galaxies.}
%\resizebox{18cm}{!}{
\setlength{\tabcolsep}{4pt}
\begin{tabular}{lccccccccc}
\hline
\hline
Galaxy    & $T$ & $D$  & Met. & $L_{[3.6]}$          & $R_{\rm eff}$ & $\Sigma_{\rm eff}$     & $R_{\rm d}$ & $\Sigma_{\rm d}$ & Ref. \\
          &     & (Mpc)&      & ($10^{9} L_{\odot}$) & (kpc)         & ($L_{\odot}$ pc$^{-2}$)& (kpc)       & ($L_{\odot}$ pc$^{-2}$) &\\
(1)       & (2) & (3)  & (4)  & (5)                  & (6)           & (7)                    & (8)         & (9) & (10)\\
\hline
NGC\,2685 & S0 & 16.1$\pm$4.8 & 1 & 30.0$\pm$0.3 & 1.92 & 1284.8 & 3.19 &  413.8 & 1\\
NGC\,2824 & S0 & 39.6$\pm$9.9 & 1 & 39.6$\pm$0.7 & 1.09 & 5209.9 & 5.00 &   84.1 & 1\\
NGC\,2859 & S0 & 26.2$\pm$6.5 & 1 & 89.1$\pm$0.8 & 1.59 & 5608.3 & 2.51 &  603.7 & 1\\
NGC\,2974 & E4 & 22.1$\pm$2.8 & 2 & 116.9$\pm$1.1& 3.02 & 1516.5 & 3.44 & 1530.5 & 2\\
NGC\,3522 & E  & 25.2$\pm$3.2 & 2 & 13.2$\pm$0.2 & 1.42 & 1039.5 & 2.47 &  254.0 & 1\\
NGC\,3626 & S0 & 22.9$\pm$2.5 & 2 & 74.6$\pm$0.7 & 2.43 & 2017.6 & 2.52 & 1891.6 & 1\\
NGC\,3838 & S0 & 23.1$\pm$5.8 & 1 & 22.5$\pm$0.2 & 1.03 & 3410.6 & 1.99 &  417.7 & 1\\
NGC\,3941 & S0 & 12.4$\pm$1.6 & 2 & 43.7$\pm$0.4 & 1.42 & 3474.0 & 1.30 & 4373.5 & 1\\
NGC\,3945 & S0 & 22.8$\pm$5.7 & 1 & 126.3$\pm$1.2& 2.50 & 3197.6 & 7.22 &  209.3 & 1\\
NGC\,3998 & S0 & 14.2$\pm$1.8 & 2 & 56.2$\pm$0.5 & 1.01 & 8807.0 & 1.46 & 2998.0 & 1\\
NGC\,4203 & S0 & 15.3$\pm$2.0 & 2 & 61.7$\pm$0.6 & 1.85 & 2093.3 & 1.85 & 2132.2 & 1\\
NGC\,4262 & S0 & 16.6$\pm$1.8 & 2 & 29.0$\pm$0.3 & 0.69 & 9480.5 & 2.90 & 43.74  & 1\\
NGC\,4278 &E1-2& 15.4$\pm$2.0 & 2 & 82.6$\pm$0.1 & 1.63 & 4884.6 & 2.57 &  992.7 & 1\\
NGC\,5582 & E  & 28.1$\pm$3.6 & 2 & 55.2$\pm$0.5 & 3.07 &  930.7 & 3.86 &  686.8 & 1\\
NGC\,6798 & S0 & 37.3$\pm$9.3 & 1 & 69.1$\pm$0.6 & 3.30 & 992.7  & 3.29 &  974.6 & 1\\
UGC\,6176 & S0 & 39.3$\pm$9.8 & 1 & 29.0$\pm$0.5 & 1.18 & 3317.6 & 1.80 & 1049.1 & 1\\
\hline
NGC\,720  & E5 & 26.1$\pm$0.0 & 2 & 234.5$\pm$0.1 & 4.5 & 1857.1 & ... & ...    & 3\\
NGC\,1332 & S0 & 24.5$\pm$0.0 & 2 & 211.0$\pm$0.1 & 2.7 & 4579.6 & 3.17& 6498.7 & 4\\
NGC\,1407 & E0 & 28.2$\pm$0.0 & 2 & 540.1$\pm$0.1 & 7.8 & 1408.7 & ... & ...    & 5\\
NGC\,1521 & E3 & 69.5$\pm$0.0 & 2 & 532.6$\pm$0.1 &10.1 & 825.7  & ... & ...    & 6\\
NGC\,4125 & E6 & 24.0$\pm$0.0 & 2 & 303.1$\pm$0.1 & 5.7 & 1448.2 & ... & ...    & 5\\
NGC\,4261 & E2 & 32.4$\pm$0.0 & 2 & 345.4$\pm$0.1 & 4.5 & 2635.3 & ... & ...    & 4\\
NGC\,4472 & E2 & 16.1$\pm$0.0 & 2 & 447.6$\pm$0.1 & 5.2 & 2587.2 & ... & ...    & 4\\
NGC\,4649 & E2 & 17.4$\pm$0.0 & 2 & 345.7$\pm$0.1 & 3.6 & 4293.7 & ... & ...    & 7\\
NGC\,6482 & E: & 57.9$\pm$0.0 & 1 & 427.9$\pm$3.9 & 4.4 & 3410.6 & ... & ...    & 5\\
\hline
\end{tabular}
%}
%\tablecomments{Column 4 gives the distance method: (1) Virgocentric infall model (as in Paper\,I), (2) surface brightness fluctuations \citep[from][]{Tully2013}. Colum 10 gives references for the kinematic or X-ray data: (1) \citet{Serra2016}, (2) \citet{Weijmans2008}, (3) \citet{Humphrey2011}, (4) \citet{Humphrey2009}, (5) \citet{Humphrey2006}, (6) \citet{Humphrey2012}, (8) \citet{Humphrey2008}.}
\label{tab:ETG}
\end{center}
\end{table*}

\section{Sample of early-type galaxies}\label{app:ETG}

Table\,\ref{tab:dSph} gives the properties of 25 ETGs. The horizontal line distinguishes between rotating ETGs from Atlas$^{\rm 3D}$ and X-ray ETGs with $Chandra$ data.\\
Column (1) gives the galaxy name.\\
Column (2) gives the Hubble type using the RC3 morphological classification \citep{RC3}.\\
Column (3) gives the adopted distance.\\
Column (4) gives the distance method: 1 = Virgocentric infall model (as in Paper\,I), 2 = surface bright fluctuations \citep[taken from][]{Tully2013}.\\
Column (5) gives the total luminosity at 3.6 $\mu$m.\\
Column (6) gives the effective radius encompassing half of the total [3.6] luminosity.\\
Column (7) gives the effective surface brightness.\\
Column (8) gives the exponential disk scale length.\\
Column (9) gives the central disk surface brightness.\\
Column (10) gives the reference for the kinematic or X-ray data: (1) \citet{Serra2016}, (2) \citet{Weijmans2008}, (3) \citet{Humphrey2011}, (4) \citet{Humphrey2009}, (5) \citet{Humphrey2006}, (6) \citet{Humphrey2012}, (8) \citet{Humphrey2008}.

\begin{table*}
\begin{center}
\caption{Catalogue of dwarf spheroidals in the Local Group}
\resizebox{18cm}{!}{
\setlength{\tabcolsep}{4pt}
\begin{tabular}{lccccccccccc}
\hline
\hline
Galaxy & $D_{\odot}$ & $D_{\rm host}$ & $\log(L_{\rm V})$  & $r_{1/2}$ & $\epsilon$ & $\sigma_{\star}$ & $N_{\star}$ & $\log(g_{\rm bar})$ & $\log(g_{\rm obs})$ & Ref.\\
       & (kpc)       & (kpc)          & ($L_{\odot}$)      & (pc)      &            & (km s$^{-1}$)    &             &  (m s$^{-2}$)       & (m s$^{-2}$) & \\
(1)         & (2)       & (3)& (4) & (5) & (6) & (7) & (8) & (9) & (10) & (11) \\
\hline
\textbf{Carina} & 105$+$6$-$6 & 107 & 5.57$\pm$0.20 & 273$\pm$45 & 0.33 & 6.60$+$1.20$-$1.20 & 774 & -12.15$+$0.27$-$0.27 & -10.81$+$0.18$-$0.18 & 1, 2, 3\\
\textbf{Draco} & 76$+$5$-$5 & 76 & 5.45$\pm$0.08 & 244$\pm$9 & 0.31 & 9.10$+$1.20$-$1.20 & 413 & -12.18$+$0.14$-$0.14 & -10.48$+$0.12$-$0.12 & 4, 5, 6\\
\textbf{Fornax} & 147$+$12$-$12 & 149 & 7.31$\pm$0.12 & 792$\pm$58 & 0.30 & 11.70$+$0.90$-$0.90 & 2483 & -11.35$+$0.17$-$0.17 & -10.77$+$0.08$-$0.08 & 1, 2, 3\\
\textbf{Leo I} & 254$+$19$-$16 & 258 & 6.74$\pm$0.12 & 298$\pm$29 & 0.21 & 9.20$+$0.40$-$0.40 & 328 & -11.06$+$0.18$-$0.18 & -10.56$+$0.06$-$0.06 & 7, 2, 8\\
\textbf{Leo II} & 233$+$15$-$15 & 236 & 5.87$\pm$0.12 & 219$\pm$52 & 0.13 & 6.60$+$0.70$-$0.70 & 171 & -11.67$+$0.26$-$0.26 & -10.71$+$0.14$-$0.14 & 9, 2, 10\\
\textbf{Sculptor} & 86$+$6$-$6 & 86 & 6.36$\pm$0.20 & 311$\pm$46 & 0.32 & 9.20$+$1.10$-$1.10 & 1365 & -11.48$+$0.26$-$0.26 & -10.58$+$0.13$-$0.13 & 11, 2, 3\\
\textbf{Sextans} & 86$+$4$-$4 & 89 & 5.64$\pm$0.20 & 748$\pm$66 & 0.35 & 7.90$+$1.30$-$1.30 & 441 & -12.96$+$0.24$-$0.24 & -11.09$+$0.15$-$0.15 & 12, 2, 3\\
Ursa Minor & 76$+$4$-$4 & 78 & 5.45$\pm$0.20 & 398$\pm$44 & 0.54 & 9.50$+$1.20$-$1.20 & 212 & -12.60$+$0.25$-$0.25 & -10.66$+$0.12$-$0.12 & 13, 2, 14\\
\textbf{Bootes I} & 66$+$3$-$3 & 64 & 4.29$\pm$0.08 & 283$\pm$7 & 0.22 & 4.60$+$0.80$-$0.60 & 65 & -13.47$+$0.14$-$0.14 & -11.14$+$0.15$-$0.12 & 15, 16, 17\\
Bootes II & 42$+$2$-$2 & 40 & 3.02$\pm$0.36 & 61$\pm$24 & 0.21 & 10.50$+$7.40$-$7.40 & 5 & -13.40$+$0.50$-$0.53 & -9.75$+$0.63$-$0.64 & 18, 5, 19\\
\textbf{Canes Venatici I} & 216$+$8$-$8 & 216 & 5.08$\pm$0.08 & 647$\pm$27 & 0.30 & 7.60$+$0.40$-$0.40 & 214 & -13.40$+$0.14$-$0.14 & -11.06$+$0.05$-$0.05 & 16, 16, 20\\
\textbf{Canes Venatici II} & 160$+$5$-$4 & 161 & 4.10$\pm$0.08 & 101$\pm$5 & 0.23 & 4.60$+$1.00$-$1.00 & 25 & -12.76$+$0.14$-$0.14 & -10.69$+$0.19$-$0.19 & 21, 16, 20\\
\textbf{Coma Berenices} & 44$+$4$-$4 & 45 & 3.46$\pm$0.24 & 79$\pm$6 & 0.36 & 4.60$+$0.80$-$0.80 & 59 & -13.19$+$0.27$-$0.27 & -10.59$+$0.16$-$0.16 & 22, 23, 20\\
Hercules & 133$+$6$-$6 & 128 & 4.42$\pm$0.16 & 175$\pm$22 & 0.67 & 3.72$+$0.91$-$0.91 & 18 & -12.92$+$0.22$-$0.22 & -11.11$+$0.22$-$0.22 & 24, 24, 25\\
Horologium I & 79$+$0$-$0 & 79 & 3.29$\pm$0.04 & 40$\pm$5 & 0.28 & 4.90$+$2.80$-$0.90 & 5 & -12.78$+$0.17$-$0.15 & -10.24$+$0.50$-$0.17 & 26, 26, 27\\
\textbf{Hydra II} & 134$+$10$-$10 & 131 & 3.87$\pm$0.12 & 88$\pm$17 & 0.01 & 4.50$+$0.00$-$4.50 & 13 & -12.88$+$0.28$-$0.19 & -10.65$+$0.12$-$0.87 & 28, 28, 29\\
\textbf{Leo IV} & 154$+$5$-$5 & 155 & 3.91$\pm$0.08 & 149$\pm$47 & 0.04 & 3.30$+$1.70$-$1.70 & 18 & -13.30$+$0.30$-$0.30 & -11.15$+$0.47$-$0.47 & 30, 16, 20\\
Leo V & 175$+$9$-$9 & 176 & 4.02$\pm$0.16 & 125$\pm$47 & 0.50 & 2.40$+$2.40$-$1.40 & 5 & -13.03$+$0.38$-$0.38 & -11.35$+$0.88$-$0.53 & 31, 31, 32\\
\textbf{Leo T} & 417$+$19$-$19 & 422 & 5.57$\pm$0.07 & 160$\pm$10 & 0.10 & 7.50$+$1.60$-$1.60 & 19 & -11.70$+$0.14$-$0.14 & -10.47$+$0.19$-$0.19 & 33, 34, 20\\
Pisces II & 196$+$15$-$15 & 195 & 3.64$\pm$0.16 & 70$\pm$18 & 0.33 & 5.40$+$3.60$-$2.40 & 7 & -12.91$+$0.29$-$0.29 & -10.39$+$0.59$-$0.40 & 35, 35, 29\\
Reticulum II & 30$+$15$-$15 & 31 & 3.01$\pm$0.04 & 42$\pm$2 & 0.59 & 3.22$+$1.64$-$0.49 & 18 & -13.10$+$0.13$-$0.12 & -10.62$+$0.49$-$0.25 & 26, 26, 27\\
Segue I & 23$+$2$-$2 & 28 & 2.54$\pm$0.32 & 28$\pm$9 & 0.48 & 3.70$+$1.40$-$1.10 & 71 & -13.22$+$0.45$-$0.42 & -10.33$+$0.36$-$0.29 & 22, 5, 36\\
\textbf{Segue II} & 35$+$2$-$2 & 41 & 2.94$\pm$0.12 & 43$\pm$6 & 0.15 & 2.60$+$0.00$-$2.60 & 25 & -13.17$+$0.20$-$0.20 & -10.81$+$0.06$-$0.87 & 37, 37, 38\\
Tucana & 887$+$49$-$49 & 883 & 5.75$\pm$0.08 & 273$\pm$52 & 0.48 & 15.80$+$4.10$-$3.10 & 17 & -11.98$+$0.21$-$0.21 & -10.05$+$0.24$-$0.19 & 39, 40, 41\\
Ursa Major I& 97$+$4$-$4 & 101 & 4.14$\pm$0.12 & 190$\pm$46 & 0.80 & 7.60$+$1.00$-$1.00 & 39 & -13.27$+$0.27$-$0.26 & -10.53$+$0.16$-$0.15 & 42, 5, 20\\
Ursa Major II & 32$+$4$-$4 & 38 & 3.54$\pm$0.20 & 123$\pm$6 & 0.50 & 6.70$+$1.40$-$1.40 & 20 & -13.49$+$0.23$-$0.23 & -10.45$+$0.19$-$0.19 & 43, 23, 20\\
Willman I & 38$+$7$-$7 & 43 & 3.01$\pm$0.28 & 25$\pm$5 & 0.47 & 4.00$+$0.80$-$0.80 & 40 & -12.63$+$0.33$-$0.36 & -10.20$+$0.20$-$0.22 & 44, 5, 45\\
\hline
\textbf{NGC 147} & 712$+$19$-$21 & 118 & 7.84$\pm$0.04 & 672$\pm$23 & 0.41 & 16.00$+$1.00$-$1.00 & 520 & -10.67$+$0.12$-$0.12 & -10.43$+$0.06$-$0.06 & 46, 40, 47\\
\textbf{NGC 185} & 620$+$18$-$19 & 181 & 7.84$\pm$0.04 & 565$\pm$7 & 0.15 & 24.00$+$1.00$-$1.00 & 442 & -10.52$+$0.12$-$0.12 & -10.00$+$0.04$-$0.04 & 46, 40, 47\\
\textbf{NGC 205} & 824$+$27$-$27 & 46 & 8.52$\pm$0.04 & 594$\pm$107 & 0.43 & 35.00$+$5.00$-$5.00 & 725 & -9.88$+$0.19$-$0.19 & -9.70$+$0.15$-$0.15 & 48, 40, 49\\
And I & 727$+$17$-$18 & 68 & 6.58$\pm$0.04 & 772$\pm$85 & 0.22 & 10.20$+$1.90$-$1.90 & 51 & -12.06$+$0.15$-$0.15 & -10.88$+$0.17$-$0.17 & 46, 50, 51\\
\textbf{And II} & 630$+$15$-$15 & 195 & 6.85$\pm$0.08 & 1355$\pm$142 & 0.20 & 9.25$+$1.10$-$1.10 & 531 & -12.27$+$0.16$-$0.16 & -11.21$+$0.11$-$0.11 & 46, 50, 52\\
And III & 723$+$24$-$18 & 86 & 5.89$\pm$0.12 & 427$\pm$43 & 0.52 & 4.70$+$1.80$-$1.80 & 43 & -12.23$+$0.18$-$0.18 & -11.30$+$0.34$-$0.34 & 46, 50, 53\\
\textbf{And V} & 742$+$22$-$21 & 113 & 5.55$\pm$0.08 & 365$\pm$57 & 0.18 & 10.50$+$1.10$-$1.10 & 85 & -12.43$+$0.19$-$0.19 & -10.53$+$0.11$-$0.11 & 46, 50, 51\\
\textbf{And VI} & 783$+$25$-$25 & 268 & 6.44$\pm$0.08 & 537$\pm$54 & 0.41 & 12.40$+$1.50$-$1.30 & 43 & -11.88$+$0.16$-$0.16 & -10.55$+$0.11$-$0.10 & 48, 50, 54\\
\textbf{And VII} & 762$+$35$-$35 & 217 & 6.98$\pm$0.12 & 965$\pm$52 & 0.13 & 13.00$+$1.00$-$1.00 & 136 & -11.85$+$0.17$-$0.17 & -10.77$+$0.07$-$0.07 & 48, 50, 51\\
\textbf{And IX} & 600$+$23$-$91 & 182 & 4.97$\pm$0.44 & 582$\pm$23 & 0.12 & 10.90$+$2.00$-$2.00 & 32 & -13.42$+$0.45$-$0.45 & -10.70$+$0.17$-$0.16 & 46, 40, 51\\
\textbf{And X} & 827$+$25$-$23 & 92 & 5.13$\pm$0.40 & 312$\pm$41 & 0.44 & 3.90$+$1.20$-$1.20 & 22 & -12.72$+$0.43$-$0.43 & -11.32$+$0.27$-$0.27 & 46, 55, 56\\
And XI & 763$+$106$-$29 & 102 & 4.70$\pm$0.48 & 210$\pm$9 & 0.24 & 7.60$+$4.00$-$2.80 & 5 & -12.80$+$0.49$-$0.49 & -10.57$+$0.46$-$0.33 & 46, 40, 54\\
And XII & 928$+$136$-$40 & 181 & 4.55$\pm$0.48 & 432$\pm$72 & 0.39 & 4.00$+$0.00$-$4.00 & 8 & -13.58$+$0.51$-$0.51 & -11.44$+$0.07$-$0.87 & 46, 40, 54\\
And XIII & 760$+$156$-$126 & 115 & 4.45$\pm$0.48 & 230$\pm$24 & 0.54 & 5.80$+$2.00$-$2.00 & 12 & -13.12$+$0.50$-$0.50 & -10.85$+$0.31$-$0.32 & 46, 40, 51\\
\textbf{And XIV} & 793$+$179$-$23 & 161 & 5.37$\pm$0.20 & 434$\pm$212 & 0.31 & 5.30$+$1.00$-$1.00 & 48 & -12.76$+$0.48$-$0.48 & -11.20$+$0.27$-$0.29 & 46, 40, 51\\
\textbf{And XV} & 626$+$35$-$79 & 174 & 5.69$\pm$0.12 & 294$\pm$12 & 0.24 & 4.00$+$1.40$-$1.40 & 29 & -12.11$+$0.17$-$0.17 & -11.28$+$0.31$-$0.31 & 46, 40, 51\\
\textbf{And XVI} & 476$+$29$-$44 & 319 & 5.53$\pm$0.12 & 164$\pm$9 & 0.29 & 5.80$+$1.10$-$0.90 & 20 & -11.76$+$0.17$-$0.17 & -10.70$+$0.17$-$0.14 & 46, 40, 57\\
\textbf{And XVII} & 727$+$25$-$39 & 67 & 5.34$\pm$0.16 & 299$\pm$31 & 0.27 & 6.50$+$3.30$-$2.70 & 16 & -12.47$+$0.21$-$0.21 & -10.86$+$0.44$-$0.36 & 46, 55, 57\\
And XIX & 821$+$148$-$32 & 116 & 5.52$\pm$0.24 & 1799$\pm$52 & 0.17 & 4.70$+$1.60$-$1.40 & 26 & -13.84$+$0.26$-$0.26 & -11.92$+$0.30$-$0.27 & 46, 58, 54\\
And XX & 741$+$52$-$42 & 128 & 4.39$\pm$0.32 & 127$\pm$36 & 0.30 & 7.10$+$3.90$-$2.50 & 4 & -12.67$+$0.45$-$0.39 & -10.42$+$0.50$-$0.32 & 46, 58, 54\\
And XXI & 827$+$25$-$23 & 135 & 5.85$\pm$0.24 & 1004$\pm$123 & 0.20 & 4.50$+$1.20$-$1.00 & 32 & -13.01$+$0.28$-$0.28 & -11.71$+$0.24$-$0.20 & 46, 59, 54\\
And XXII & 920$+$139$-$32 & 275 & 4.66$\pm$0.32 & 223$\pm$60 & 0.56 & 2.80$+$1.90$-$1.40 & 10 & -12.89$+$0.41$-$0.41 & -11.47$+$0.60$-$0.45 & 46, 59, 54\\
\textbf{And XXIII} & 748$+$21$-$31 & 127 & 6.00$\pm$0.20 & 1034$\pm$97 & 0.40 & 7.10$+$1.00$-$1.00 & 42 & -12.88$+$0.24$-$0.24 & -11.32$+$0.13$-$0.13 & 46, 60, 54\\
\textbf{And XXIV} & 898$+$42$-$28 & 169 & 5.32$\pm$0.20 & 633$\pm$52 & 0.25 & 7.30$+$0.00$-$7.30 & 12 & -13.14$+$0.24$-$0.24 & -11.09$+$0.04$-$0.87 & 46, 60, 54\\
And XXV & 736$+$69$-$23 & 90 & 5.75$\pm$0.20 & 742$\pm$70 & 0.25 & 3.00$+$1.20$-$1.10 & 26 & -12.85$+$0.24$-$0.24 & -11.93$+$0.35$-$0.32 & 46, 60, 54\\
And XXVI & 754$+$164$-$218 & 103 & 4.77$\pm$0.20 & 253$\pm$30 & 0.25 & 8.60$+$2.80$-$2.20 & 6 & -12.90$+$0.25$-$0.25 & -10.55$+$0.31$-$0.25 & 46, 60, 54\\
And XXVII & 1214$+$474$-$42 & 482 & 5.45$\pm$0.20 & 679$\pm$253 & 0.40 & 14.80$+$4.30$-$3.10 & 8 & -13.07$+$0.40$-$0.40 & -10.50$+$0.30$-$0.30 & 46, 60, 54\\
\textbf{And XXVIII} & 811$+$48$-$48 & 384 & 5.39$\pm$0.16 & 285$\pm$12 & 0.43 & 4.90$+$1.60$-$1.60 & 18 & -12.37$+$0.20$-$0.20 & -11.09$+$0.29$-$0.29 & 61, 61, 62\\
\textbf{And XXIX} & 829$+$42$-$42 & 198 & 5.33$\pm$0.12 & 377$\pm$30 & 0.29 & 5.70$+$1.20$-$1.20 & 24 & -12.68$+$0.18$-$0.18 & -11.08$+$0.19$-$0.19 & 61, 61, 62\\
And XXX & 681$+$78$-$32 & 145 & 5.20$\pm$0.00 & 306$\pm$45 & 0.38 & 11.80$+$7.70$-$4.70 & 8 & -12.63$+$0.17$-$0.17 & -10.35$+$0.57$-$0.36 & 46, 63, 54\\
\textbf{And XXXI} & 756$+$28$-$44 & 262 & 6.61$\pm$0.28 & 1231$\pm$132 & 0.43 & 10.30$+$0.90$-$0.90 & 126 & -12.43$+$0.31$-$0.32 & -11.08$+$0.09$-$0.09 & 64, 64, 65\\
And XXXII & 772$+$56$-$61 & 140 & 6.83$\pm$0.28 & 1376$\pm$340 & 0.50 & 8.40$+$0.60$-$0.60 & 212 & -12.31$+$0.37$-$0.36 & -11.30$+$0.13$-$0.12 & 64, 64, 65\\
\textbf{Cetus} & 779$+$43$-$43 & 688 & 6.44$\pm$0.08 & 791$\pm$75 & 0.33 & 8.30$+$1.00$-$1.00 & 11 & -12.22$+$0.16$-$0.16 & -11.07$+$0.11$-$0.11 & 39, 50, 66\\
\textbf{Perseus I} & 785$+$65$-$65 & 351 & 6.04$\pm$0.28 & 391$\pm$143 & 0.43 & 4.20$+$3.60$-$4.20 & 12 & -12.00$+$0.43$-$0.45 & -11.36$+$0.76$-$0.88 & 67, 67, 65\\
Pisces I & 769$+$24$-$24 & 268 & 5.98$\pm$0.04 & 560$\pm$53 & 0.20 & 7.90$+$5.30$-$2.90 & 4 & -12.37$+$0.14$-$0.14 & -10.97$+$0.58$-$0.32 & 48, 40, 68\\
\hline
\end{tabular}
}
\tablecomments{The total luminosity of Leo T includes half of the gas mass, giving the actual total baryonic mass for $\Upsilon_{\rm V} = 2$ $M_{\odot}/L_{\odot}$. Galaxies in bold are included in our high-quality sample (see Sect.\,\ref{sec:dSphs}).}
\label{tab:dSph}
\end{center}
\end{table*}

\section{Catalogue of dwarf spheroidals}\label{sec:Catalogue}

Table\,\ref{tab:dSph} gives the properties of 62 dSphs in the Local Group. The horizontal line distinguishes between MW and M31 satellites (whichever is closer). Objects like Tucana and Cetus are relatively isolated, but they are considered as ``satellites'' for simplicity. Galaxies are listed in alphabetic order, but we start with the ``classic'' satellites and follow with the ultrafaint ones. Bolded galaxies represent our high-quality sample (see Sect.\,\ref{sec:dSphs}). \\
Column (1) gives the galaxy name.\\
Column (2) gives the heliocentric distance derived from the tip of the red giant branch method or variable stars.\\
Column (3) gives the distance to the host galaxy.\\
Column (4) gives the total luminosity in $V$-band adopting a solar absolute magnitude of 4.83.\\
Column (5) gives the deprojected 3D half-light radii ($r_{1/2} = 4/3 R_{1/2}$, see \citealt{Wolf2010}).\\
Column (6) gives the observed ellipticity.\\
Column (7) gives the observed mean velocity dispersion.\\
Column (8) gives the number of stars used to estimate the mean velocity dispersion $\sigma_{\star}$.\\
Column (9) gives the baryonic acceleration (Eq.\,\ref{Eq:dSph1}).\\
Column (10) gives the observed acceleration (Eq.\,\ref{Eq:dSph2}).\\
Column (11) gives references for distances, structural parameters, and velocity dispersions, respectively. They are coded as follows:
1:~\citet{2009AJ....138..459P}, 2:~\citet{1995MNRAS.277.1354I}, 3:~\citet{Walker2009b}, 4:~\citet{2004AJ....127..861B}, 5:~\citet{2008ApJ...684.1075M}, 6:~\citet{2007ApJ...667L..53W}, 7:~\citet{2004MNRAS.354..708B}, 8:~\citet{2008ApJ...675..201M}, 9:~\citet{2005MNRAS.360..185B}, 10:~\citet{2007AJ....134..566K}, 11:~\citet{2008AJ....135.1993P}, 12:~\citet{2009ApJ...703..692L}, 13:~\citet{2002AJ....123.3199C}, 14:~\citet{2009ApJ...704.1274W}, 15:~\citet{2006ApJ...653L.109D}, 16:~\citet{2012ApJ...744...96O}, 17:~\citet{2011ApJ...736..146K}, 18:~\citet{2008ApJ...688..245W}, 19:~\citet{2009ApJ...690..453K}, 20:~\citet{2007ApJ...670..313S}, 21:~\citet{2008ApJ...675L..73G}, 22:~\citet{2007ApJ...654..897B}, 23:~\citet{2010AJ....140..138M}, 24:~\citet{2009ApJ...704..898S}, 25:~\citet{2009A&A...506.1147A}, 26:~\citet{2015ApJ...805..130K}, 27:~\citet{2015arXiv150407916K}, 28:~\citet{2015ApJ...804L...5M}, 29:~\citet{2015arXiv150601021K}, 30:~\citet{2009ApJ...699L.125M}, 31:~\citet{2010ApJ...710.1664D}, 32:~\citet{2009ApJ...694L.144W}, 33:~\citet{2007ApJ...656L..13I}, 34:~\citet{2008ApJ...680.1112D}, 35:~\citet{2012ApJ...756...79S}, 36:~\citet{2011ApJ...733...46S}, 37:~\citet{2009MNRAS.397.1748B}, 38:~\citet{2013ApJ...770...16K}, 39:~\citet{2009ApJ...699.1742B}, 40:~\citet{2012AJ....144....4M}, 41:~\citet{2009A&A...499..121F}, 42:~\citet{2008A&A...487..103O}, 43:~\citet{2006ApJ...650L..41Z}, 44:~\citet{2006astro.ph..3486W}, 45:~\citet{2011AJ....142..128W}, 46:~\citet{2012ApJ...758...11C}, 47:~\citet{2010ApJ...711..361G}, 48:~\citet{2005MNRAS.356..979M}, 49:~\citet{2006AJ....131..332G}, 50:~\citet{2006MNRAS.365.1263M}, 51:~\citet{2012ApJ...752...45T}, 52:~\citet{2012ApJ...758..124H}, 53:~\citet{2007ApJ...662L..79C}, 54:~\citet{2013ApJ...768..172C}, 55:~\citet{2011ApJ...729...23B}, 56:~\citet{2009ApJ...705.1043K}, 57:~\citet{2015ApJ...799L..13C}, 58:~\citet{2008ApJ...688.1009M}, 59:~\citet{2009ApJ...705..758M}, 60:~\citet{2011ApJ...732...76R}, 61:~\citet{2015ApJ...806..230S}, 62:~\citet{2013ApJ...768...50T}, 63:~\citet{2015MNRAS.450.1409S}, 64:~\citet{2013ApJ...772...15M}, 65:~\citet{2014ApJ...793L..14M}, 66:~\citet{2014MNRAS.439.1015K}, 67:~\citet{2013ApJ...779L..10M}, 68:~\citet{1999PASP..111..306C}

\bibliography{RAR}

\end{document}